\documentclass[usenatbib]{mnras}

\usepackage[T1]{fontenc}
\usepackage{ae,aecompl}

\usepackage{graphicx}	% Including figure files
\usepackage{amsmath}	% Advanced maths commands
\usepackage{mwe}
\usepackage{newtxtext,newtxmath}

\title[GC properties and multiple stellar populations]{Linking globular cluster structural parameters and their evolution: multiple stellar populations}

\author[Mastrobuono-Battisti \& Perets]{
Alessandra Mastrobuono-Battisti$^{1,2}$\thanks{mastrobuono@mpia.de}
and
Hagai B. Perets$^{3,4}$\\
$^{1}$Department of Astronomy and Theoretical Physics, Lund Observatory, Box 43, SE--221 00, Lund, Sweden\\
$^{2}$Max Planck Institute for Astronomy, K\"onigstuhl 17, D69117, Heidelberg, Germany\\
$^{3}$Physics department, Technion - Israel Institue of Technology, Haifa 3200003, Israel\\
$^{4}$Cahill Center for Astrophysics, Division of Physics, Mathematics and Astronomy, California Institute of Technology, Pasadena, CA 91125
}

\date{Accepted XXX. Received YYY; in original form ZZZ}

\pubyear{2020}

\begin{document}
\label{firstpage}
\pagerange{\pageref{firstpage}--\pageref{lastpage}}
\maketitle
\begin{abstract}
Globular clusters (GCs) are known to host multiple stellar populations showing chemical anomalies in the content of light elements. The origin of such anomalies observed in Galactic GCs is still debated. Here we analyse data compiled from the Hubble Space Telescope, ground-based  surveys and Gaia DR2 and explore relationships between the structural properties of GCs and the fraction of second population (2P) stars. Given the correlations we find, we conclude that the main factor driving the formation/evolution of 2P stars is the cluster mass. The existing strong correlations between the 2P fraction and the rotational velocity and concentration parameter could derive from their correlation with the cluster mass. Furthermore, we observe that increasing cluster escape velocity corresponds to higher 2P fractions. Each of the correlations found is bimodal, with a different behaviour detected for low and high mass (or escape velocity) clusters. These correlations could be consistent with an initial formation of more centrally concentrated 2P stars in deeper cluster potentials, followed by a long-term tidal stripping of stars from clusters outskirts. The latter are dominated by the more extended distributed first population (1P) stars, and therefore stronger tidal stripping would preferentially deplete the 1P population, raising the cluster 2P fraction. This also suggests a tighter distribution of initial 2P fractions than observed today. In addition, higher escape velocities allow better retention of low-velocity material ejected from 1P stars, providing an alternative/additional origin for the observed differences and the distributions of 2P fractions amongst GCs.
\end{abstract}
% Don't make up new ones
\begin{keywords}
Galaxy: globular clusters -- Galaxy: kinematics and dynamics -- Galaxy: stellar content
\end{keywords}

%%%%%%%%%%%%%%%%%%%%%%%%%%%%%%%%%%%%%%%%%%%%%%%%%%

%%%%%%%%%%%%%%%%% BODY OF PAPER %%%%%%%%%%%%%%%%%%

\section{Introduction}	
For decades globular clusters (GCs) have been thought to represent the perfect example of simple stellar populations.
All the stars in each GC were thought to have formed concurrently and with the same chemical composition. Since leftover gas could be ejected due to supernovae winds, 
no further star formation episodes were considered possible. 
However, photometric and spectroscopic studies collected over the last decade \citep[see e.g.][]{PBA07, CBG09} have progressively revealed a different picture, showing that most, if not all, Galactic GCs host multiple stellar populations with different
light elements content. While the first, pristine, stellar population has a chemical composition similar to halo field stars with similar metallicity, the stars belonging to the second stellar population
present numerous chemical anomalies, including anti-correlations between the content of different elements \citep[e.g. C-N, Na-O, and in some cases Mg-Al,][]{Gra12, Gratton19}. These anomalies seem to be ubiquitous, pointing toward a possible common origin of multiple populations in all Galactic GCs. Similar anomalies have been observed in stellar clusters of different ages in the LMC and SMC, making the puzzle even more complex to solve \citep[see][for a review on the topic]{BL18}.

Several scenarios have been proposed to explain the existence of the observed multiple populations, but no clear solution has been yet found \citep[see][for a summary of the scenarios and for their caveats]{Ren15, BL18, Gratton19}.
One possibility is that GCs experienced two or more star formation episodes, in which second population (2P) stars formed 
from processed gas lost by massive first population (1P) stars, and/or accreted external gas. 

In the so-called self-enrichment scenarios, in which the gas is provided by 1P stars, the nature of the possible donors is constrained by the chemical properties of the 2P stars. Different potential sources of the gas have been identified \citep[e.g. AGBs, fast-rotating massive stars][] {Ve01, De07, BR18}. 
Alternative scenarios predict that the anomalous stars form as a consequence of processes that can only happen
in primordial GCs \citep[e.g. accretion of processed material onto the zero-age main sequence stellar disc, formation and evolution of a central supermassive star][]{BA13, GCK18,Wang20}. 

The later dynamical evolution of GCs can affect the properties of the GCs and their 1P/2P stars on longer timescales, potentially  leaving observable signatures on the internal structure of the cluster. The amplitude of the left-over signatures will depend on the amount of 2P stars that form compared to the 1P stars, leading to relationships between internal cluster properties and the fraction of 2P stars.\\
For example,  the self-enrichment scenarios predict that the 2P stars form centrally concentrated. According to these models, the higher concentration should be still observed today in clusters that are not fully relaxed \citep{MBP13, Vesperini13, HB15, MBP16, TV19, Vesperini21}. Centrally concentrated 2P stars have been indeed observed in a number of clusters  \citep[see e.g.][]{NF79, MPB12, RH13, 2CPJ14,BV15,CHB17,Da19, Da20, Kamann20, Szigeti21}.  The observed differences are milder for more dynamically evolved systems \citep{Da19}.  
%If the relaxation time of a cluster is long enough these signatures could be still visible \citep{MBP13,HB15,MBP16,CHB17}. 
Moreover, simulations predict kinematic differences between the two populations, including differential rotation and discrepancies in the velocity dispersions \citep{Vesperini13, MBP13,HB15, MBP16, TV19}. This kind of differences have been observed in Galactic GCs  \citep{RH13, BV15, CHB17, LJ17, BLB18, LJ18, DM18, MM18, CM19, Kamann20}, however, the strength and variety of the signatures hamper the interpretation of the results.\\
\cite{Car10} used principal component analysis and multivariate relations to study the phase of transition between 1P and 2P, finding that the main parameters driving the chemical signatures found in GCs are the metallicity, mass, and cluster age \citep[see also][]{LG19, Martocchia18, Ma19}.  \cite{Car19} extended these results, identifying the mass and the concentration as the two most important parameters determining the extent of the chemical anomalies observed in Galactic GCs.\\
%%%%aggiungere Carretta 2019 qui e anche nelle discussioni!!!
More recently, \cite{Mi19} used the `chromosome maps' of 59 Galactic GCs and of 11 GCs of both Magellanic Clouds  to compare the multiple populations phenomenon in different environments. In this way they found correlations between the 2P properties and the internal as well as orbital parameters of the clusters. In particular, the fraction of 1P stars exhibits a strong anti-correlation with the mass of the 2P and with the initial cluster mass. They also found a mild correlation between the 2P and current cluster mass.  
The 1P fraction is correlated with the total GC luminosity and clusters with larger peri-galactic radii show larger helium variations. In addition, Galactic and extra-galactic clusters seem to follow similar relationships.

 We note that in self-accreting scenarios, material lost by 1P stars could be lost if ejected at velocities higher than the escape velocity of the cluster.  For example, mass lost through winds of evolved stars typically have velocities of 10-30 km/s, comparable with the escape velocities of clusters. Clusters with low escape velocities might therefore retain only a fraction of the mass lost through winds, while more massive and concentrated clusters with higher escape velocities would retain more. In fact, if the escape velocity from the cluster is higher than the fastest winds from evolved stars, all the mass lost through winds will be retained, and one might expect all clusters above this threshold to have comparable 2P fractions, independent of their mass.

In this work we carry out a systematic search of correlations between different internal cluster parameters, complementing previous works and extending them to the analysis of the correlations with other internal properties.
We use data collected from the literature to assess the possible relationships existing between cluster internal properties and the fraction of 2P stars. In Section \ref{sec:mm} we introduce the different data-sets that we employed in our study, in Section \ref{sec:res} we present the results of our analysis and in Section \ref{sec:con} we discuss the results and draw our conclusions.

\section{Data-sets and analysis}\label{sec:mm}
To study the correlations between internal GC parameters focusing on the presence of 2P  stars, we collected data from different sources.
The 2P number fractions, to which we refer to quantify the 2P phenomenon, are from \cite{MI17}.
We used the \cite[][2010 edition]{H96} catalogue for most of the structural parameters (metallicity, ellipticity, concentration, core radius and half-mass relaxation time).
The masses of the clusters have been obtained from \cite{GN97}.  As a further test, we have also used the masses from \cite{BH18}. %The clusters' Jacobi radii are obtained from \cite{BP10}.
The cluster ages are taken from  the most recent homogeneous compilation in the literature, presented by \cite{VDB13}. \\
Gaia DR2 provided detailed information on their radial velocities of stars in the outskirts of clusters and allowed the estimate of the clusters orbital parameters, including the radial period, $T_r$ \citep{Helmi18}. Proper motions have a larger coverage of clusters
since photometric measures are less affected by the crowding. As an estimate of the 3D rotational velocity, we used the amplitude of the rotation estimated by \cite{SBH19} from Gaia DR2 proper motions and literature line-of-sight velocities of Galactic GCs. We also used the rotation on the plane of the sky evaluated by \cite{BP18} using the same Gaia DR2 data as another a proxy of the cluster internal rotation.
%As a comparison we also used the internal parameters provided by \cite{BH18} and \cite{BA19}.
%All the data and their sources are summarised in Table XX.

We quantified the relationships between the internal properties of GCs by evaluating the Pearson correlation coefficients and probabilities. These indicators are, however, strongly affected by the presence of outliers and only check for the presence of linear correlations. Hence, we also calculated the Spearman and Kendal coefficients and probabilities, which attribute less weight to the outliers and can detect non-linear correlations. While the coefficients represent the strength and direction of the linear correlation between the two variables taken into account, we consider a probability smaller than the significance level of $0.05$ to indicate the rejection of the hypothesis that no correlation exists between the two quantities. All the tests have been run for the whole sample of clusters and for clusters less  or more massive than $10^{5.5}~M_\odot$ separately. The results of the analysis are presented in the next section. \\
In Appendix \ref{sec:app1} the same tests are repeated considering metal rich  ($[Fe/H]>-1$)  and metal poor ($[Fe/H]\leq -1$)  GCs separately, as well as for the entire sample of GCs.   and \ref{sec:app2} we detail the results obtained using the projected rotational velocity instead of the 3D one. In Appendix \ref{app:age} we discuss the age correlations obtained using different data-sets available in literature \citep{Gratton10,FO10,Do10} and in Appendix \ref{sec:app_bin_frac} we discuss the existence of a correlation between the second generation fraction and the cluster binary fraction.
 
%%%%%%%%%%%%%%%%%%%%%%%
\begin{table*}\centering
\small
\begin{tabular}{ccccccccc}
\hline
Par$_{1}$ & Par$_{2}$ & Stat & PCC & PP & PCC$_{LM}$ & PP$_{LM}$ &PCC$_{HM}$ & PP$_{HM}$\tabularnewline
\hline 
$F_{2}$ & $M$ & P &0.76 & $3.2\times10^{-11}$  &  0.50 & $0.0056$ &  0.64 & $5.3\times10^{-4}$ \tabularnewline
        &     & S &0.77 &$1.5\times10^{-11}$ & 0.54&$0.0023$& 0.62  & $9.5\times10^{-4}$  \tabularnewline
        &     & K &0.58 &$4.9\times10^{-10}$ & 0.38 &$0.0038$& 0.45  & 0.0020  \tabularnewline [0.1cm]
  
 $F_{2}$ & $c$  & P &0.50& $2.9\times10^{-4}$ &0.37& $0.063$ &0.025& $0.91$\tabularnewline
 & & S & 0.55  & $4.5\times10^{-5}$ & 0.48& $0.013$ & 0.10 & $0.64$\tabularnewline
 &  & K & 0.40  & $6.2\times10^{-5}$ & 0.40& $0.0092$ & 0.56 & $0.73$\tabularnewline [0.1cm]
        
$F_{2}$ & $|v_{t}|$ & P &0.46& $0.0043$ &  0.092 & $0.73$ & 0.34 &$0.14$\tabularnewline
 &  &  S &0.54 & $8.7\times10^{-4}$ & $0.19$ & $0.47$ & 0.41 & 0.071\tabularnewline
 &  & K &  0.38 &$8.0\times10^{-4}$ & $0.12$ & $0.56$ & 0.32 & 0.055\tabularnewline [0.1cm]

$F_{2}$ & $|v_{A}|$ & P &  0.47& $0.0016$ & $-0.25$ & 0.32&  0.38 & $0.066$\tabularnewline
 &  & S &0.43& $0.0042$ & $-0.22$ & $0.38$ &$0.32$ & 0.12  \tabularnewline
 &  & K &0.30& $0.0052$ & $-0.18$ & $0.33$ &0.21 & 0.16  \tabularnewline [0.1cm]

 $F_{2}$ & $V/\sigma$ & P &0.28& $0.11$ & 0.12 &$ 0.67$ & 0.13 & $0.60$\tabularnewline
 & & S & $0.33$  & $ 0.056$ & $-0.014$& 0.96 &0.21 & 0.38 \tabularnewline
 &  & K & $0.21$ & $ 0.083$ & $-0.017$& 0.96 &0.17 & 0.34 \tabularnewline [0.1cm]
 
 $F_{2}$ & $\varepsilon$ & P &0.082& $0.57$ & $-0.12$ & $0.56$ & 0.52 &$0.0073$\tabularnewline
 &  & S & $0.079$ & $0.58$ & $-0.22$ & 0.28 & 0.49 & 0.012\tabularnewline
 &  & K & $0.058$ & $0.57$ & $-0.15$ & 0.32 & 0.36 & 0.015\tabularnewline [0.1cm]
 
$F_{2}$ & $age$ &  P &$3.5\times10^{-3}$ & $0.98$ &0.24& $0.23$ &$-0.28$& $0.21$\tabularnewline
 &  & S& $-0.049$  & 0.74 & 0.25 & 0.22 & $-0.36$ &  0.10 \tabularnewline
 & & K & $ -0.039$  & 0.71 & 0.18 & 0.22 &$-0.28$ &  0.081 \tabularnewline [0.1cm]

$F_{2}$ & $age/t_{\rm rh}$ & P  & $-0.31$ & $0.031$ &$-0.15$& $ 0.44$ & $-0.33$ & $0.13$\tabularnewline
 &  & S &  $-0.16$ & $0.26$ & $0.14$ & 0.48 & $-0.37$ & 0.092\tabularnewline
 &  & K & $-0.11$ & $0.27$ & $0.10$ & 0.47 &$ -0.25$ & 0.11\tabularnewline [0.1cm]

$F_{2}$ & $\log(t_{\rm rh})$ & P & 0.25 & $0.071$ & $-0.068$ & $0.73$ &0.47& $0.018$\tabularnewline
 &  & S &0.14& $0.31$ & $-0.18$ & 0.35 & $0.40$ &0.049\tabularnewline
 &  & K &$0.093$& $0.32$ & $-0.13$ & 0.33& $0.28$ &0.055\tabularnewline [0.1cm]

$F_2$ & $age/T_r$ & P & $2.7\times10^{-3}$ & $0.88$ & $0.38$ & $0.091$ & $-0.36$ & $0.19$ \tabularnewline 
%&  &  &  $(0.070)$ & $(0.66)$&  $(0.31)$ & $(0.15)$&  $(-0.17)$ & $(0.50)$
 &  & S & $3.9\times10^{-3}$ & $0.98$ & $0.45$ & $0.043$ & $-0.24$ & $0.38$  \tabularnewline
 %&  &  &  $(0.036)$ & $(0.82)$&  $(0.38)$ & $(0.077)$&  $(-0.17)$ & $(0.50)$
 &  & K & $6.3\times10^{-3}$ & $0.97$ & $0.34$ & $0.031$ & $-0.16$ & $0.44$ \tabularnewline
 % &  &  &  $(0.022)$ & $(0.85)$&  $(0.26)$ & $(0.091)$&  $(-0.085)$ & $(0.65)$
\tabularnewline 
\hline 
\end{tabular}
\caption{Statistical correlation coefficients (PCC) and probabilities (PP) between $F_2$ and several GC internal parameters. The statistical parameters are reported for the entire sample of GCs as well as for clusters less massive (LM) and more massive (HM) than $10^{5.5}~M_\odot$. P stands for Pearson, S for Spearman and K for Kendall. 
}\label{tab:tab1}
\end{table*}
%%%%%%%%%%%%%%%%%%%%%%%

\begin{table*}\centering
\small
\begin{tabular}{ccccccccc}
\hline
Par$_{1}$ & Par$_{2}$ & Stat & PCC & PP & PCC$_{LM}$ & PP$_{LM}$ &PCC$_{HM}$ & PP$_{HM}$\tabularnewline
\hline 

$|v_{A}|$ & $M$ & P &0.47& $1.5\times10^{-4}$& 0.068 & $0.74$ &0.41& $0.015$ \tabularnewline
 & & S &0.51& $3.1\times10^{-5}$& $-0.088$& $0.67$ & $0.48$ & 0.0032 \tabularnewline
 & & K &$0.34$& $8.2\times10^{-5}$& $-0.074$& $0.61$ & $0.32$ & 0.0076 \tabularnewline [0.1cm]

$|v_{A}|$ &  $ c$  & P &0.43 & $0.0011$ & 0.29 & $0.18$ & 0.32& $0.069$\tabularnewline
 &    & S & $0.39$ & $0.0026$ & $0.094$ & 0.67 & 0.20 &  0.28 \tabularnewline
 &    & K & $0.27$ & $0.0034$ & $0.044$ & 0.79 & 0.13 &  0.30 \tabularnewline [0.1cm]

$|v_{A}|$ & $age$  & P &0.083& $0.61$ & 0.16 & $0.95$ &0.18& $0.43$\tabularnewline
 &   & S & $0.13$ & $0.41$ & $0.071$ & 0.77 & 0.23 & 0.31 \tabularnewline
 &   & K & $0.077$ & $0.50$ & $0.085$ & 0.65 & 0.13 & 0.44 \tabularnewline [0.1cm]

$|v_{A}|$ & $age/t_{\rm rh}$ & P  & $-0.33$& $0.038$ & $-0.21$& $0.38$ &  $-0.42$ & $0.055$\tabularnewline
 &  & S & $-0.30$& $0.060$ &$ -0.072$& $0.77$ &  $-0.44$ & $0.042$ \tabularnewline
 &  & K &$-0.22$ &0.047 &  $-0.041$ &0.84& $-0.33$ & 0.034 \tabularnewline [0.1cm]

$|v_{A}|$ & $\varepsilon$  & P & 0.21& $0.11$ &  0.079 & $0.73$ & 0.26 & $0.13$\tabularnewline
 &   & S & $0.26$ & $0.048$ & $0.26$ & 0.24 & 0.30 & 0.083 \tabularnewline
 &   & K & $0.19$ & $0.045$ & $0.20$ & 0.21 & 0.22& 0.074 \tabularnewline [0.1cm]

$M$ & $V/\sigma$ & P  &0.16& $0.28$ & 0.34& $0.13$ &$-0.065$& $0.75$\tabularnewline
 &  & S & $ 0.34$ & $ 0.019$ & $0.29$ & 0.20 & 0.079 & 0.70\tabularnewline
 &  & K & $ 0.21$ & $ 0.037$ & 0.18&  $0.26$ & 0.038 & 0.80 \tabularnewline [0.1cm]

$M$ &  $c$   & P & $ 0.30$ & $0.0021$ & 0.25&  $0.049$ &0.10 & 0.53\tabularnewline
 &     & S & $ 0.47$ & $5.3\times10^{-7}$ & 0.37& $0.0028$&  0.23 & 0.16\tabularnewline
 &     & K & $ 0.34$ & $ 3.6\times10^{-7}$ & 0.26&  $0.0027$ &0.16 & 0.14\tabularnewline [0.1cm]

$M$ &  $age$   & P & $ 0.010$ & $0.94$ & 0.34&  $0.059$ &-0.031 & 0.88\tabularnewline
 &     & S &  $ 0.030$ & $0.83$ & 0.23&  $0.22$ &0.89 & 0.72\tabularnewline
 &     & K & $8.8\times10^{-3}$ & $0.93$ & 0.16&  $0.23$ &0.00 & 1.00\tabularnewline [0.1cm]

$M$ & $age/t_{\rm rh}$ & P  &$-0.32$ &0.019 &  $-0.15$ &0.44 &  -0.65& $8.3\times10^{-4}$ \tabularnewline
 &   & S &$-0.27$ &0.047 &  0.083 &0.66 &  $-0.75$& $3.5\times10^{-5}$ \tabularnewline
 &   & K &$-0.20$ &0.036 &  0.058 &0.66 &  $-0.56$& $2.4\times10^{-4}$ \tabularnewline [0.1cm]

$M$ & $\varepsilon$  & P & $ 0.17$ & $ 0.096$ & 0.065&  $0.64$ &0.28 & 0.064 \tabularnewline 
 &  & S & $ 0.17$ & $ 0.10$ & 0.15&  $0.29$ &0.16 & 0.31 \tabularnewline 
 &  & K & $0.13$ & $0.076$ & $0.11$ &  $0.27$ &0.13 & 0.24 \tabularnewline [0.1cm]
\hline 
\end{tabular}
\caption{Statistical correlation coefficients (PCC) and probabilities (PP) between the 3D rotational velocity amplitude, $v_A$, age, mass and several GC internal cluster parameters. Parameters are reported for the entire sample of GCs as well as for clusters less massive (LM) and more massive (HM) than $10^{5.5}~M_\odot$. P stands for Pearson, S for Spearman and K for Kendall. }
\label{tab:tab2}
\end{table*}
%%%%%%%%%%%%%%%%%%%%%%%%%%%%%%%%
\begin{figure*}\centering
\includegraphics[width=0.85\textwidth]{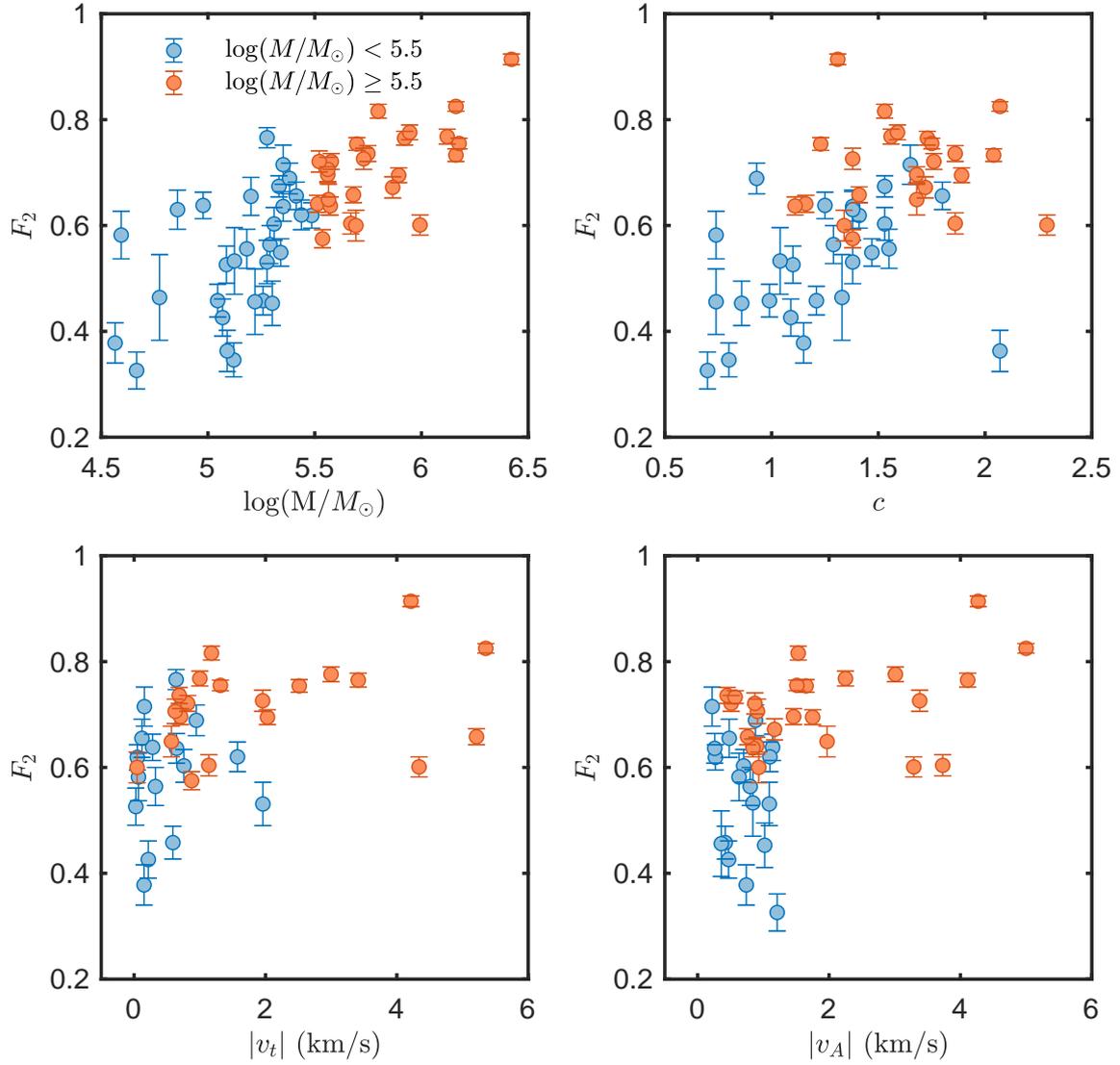}
\caption{ The second generation number fraction, $F_2$, for the available cluster sample is plotted against different GC internal parameters. From top left to bottom right, $F_2$ is plotted against the logarithm of the current cluster masse,  the cluster concentration, the rotational velocity of the clusters on the plane of the sky and the 3D rotational velocity. 
Orange are clusters with masses larger than $10^{5.5}~M_\odot$ and light blue are clusters with masses smaller than the same threshold.
}\label{fig:strongcorr}
\end{figure*}
%%%%%%%%%%%%%%%%%%%%
\begin{figure*}\centering
\includegraphics[width=0.85\textwidth]{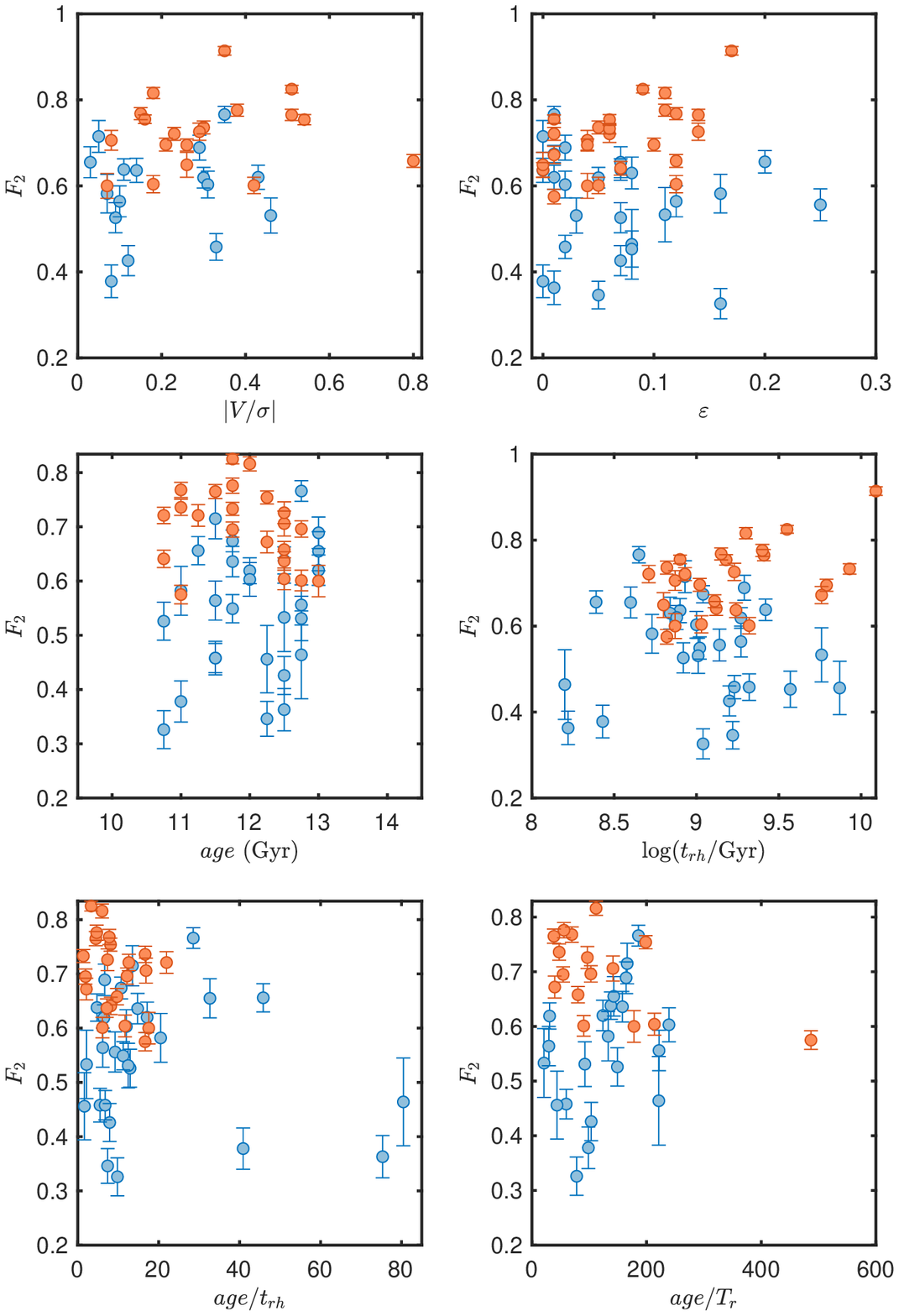}
\caption{ The second generation stellar number fraction, $F_2$, is plotted against the cluster $|V/\sigma|$ parameter (top left panel), the cluster ellipticity (top right panel), the cluster age  (middle left panel), the logarithm of the half-mass relaxation time (middle right panel),  the dynamical age (bottom left panel) and the age over period (bottom rigth panel). 
Orange are clusters with masses larger than $10^{5.5}~M_\odot$ and light blue are clusters with masses smaller than the same threshold.}\label{fig:other_corr}
\end{figure*} 
%%%%%%%%%%%%%%%%%%%%%%%%%%%%%%%

\section{Results}\label{sec:res}
Currently, none of the proposed formation scenarios is able to reproduce all the chemical signatures observed for the stars belonging to the second population in each cluster \citep{Ren15, BL18, Gratton19}.
In the following, we present the resulting relations between the cluster structural and kinematic properties as well as the fractions of 2P stars.
The parameter space that needs to be explored is extremely large. Therefore, we restricted our search to the most promising structural parameters, like the cluster masses, ages, rotation and concentrations.
As these results can come from existing correlations between the 2P fraction and the internal properties of the clusters, they  might shed light on the formation scenario. \\
 For the first time, we conduct a systematic search which considers all the relevant structural parameters for all the observed Galactic GCs and that adopts the second generation number fraction as the only reference to quantify the second population phenomenon.

\begin{figure*}\centering
\includegraphics[width=0.85\textwidth]{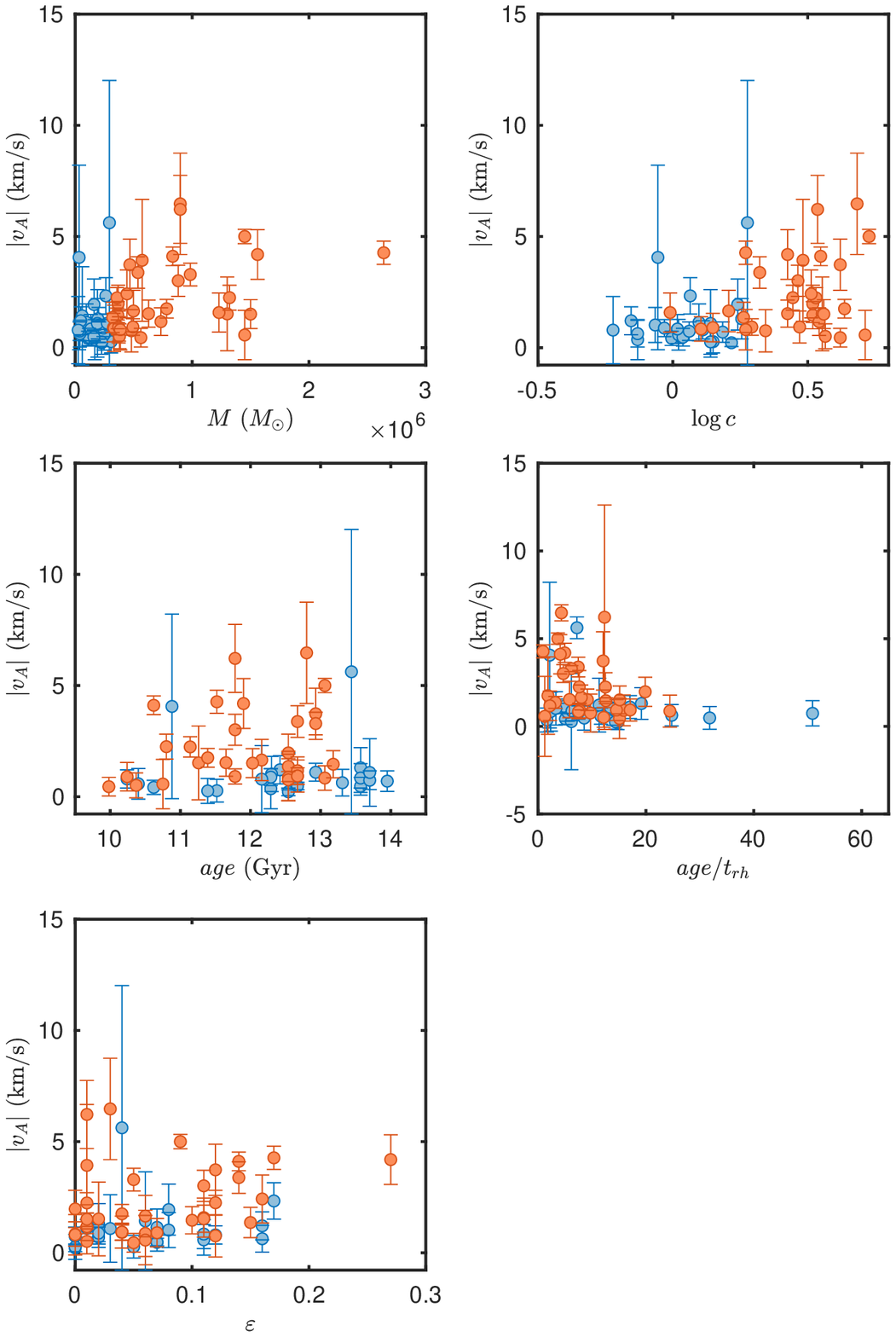}
\caption{The 3D amplitude of the rotational velocity for the available cluster sample, $|v_A|$, is plotted against the mass of the clusters (top left panel), the logarithm of the concentration parameter top right panel), the cluster age (middle left panel), the dynamical age of the clusters (middle right panel) and the cluster ellipticity (bottom left panel). 
Orange are clusters with masses larger than $10^{5.5}~M_\odot$ and light blue are clusters with masses smaller than the same threshold.}
\label{fig:vtcorr}
\end{figure*}

\begin{figure*}\centering
\includegraphics[width=0.85\textwidth]{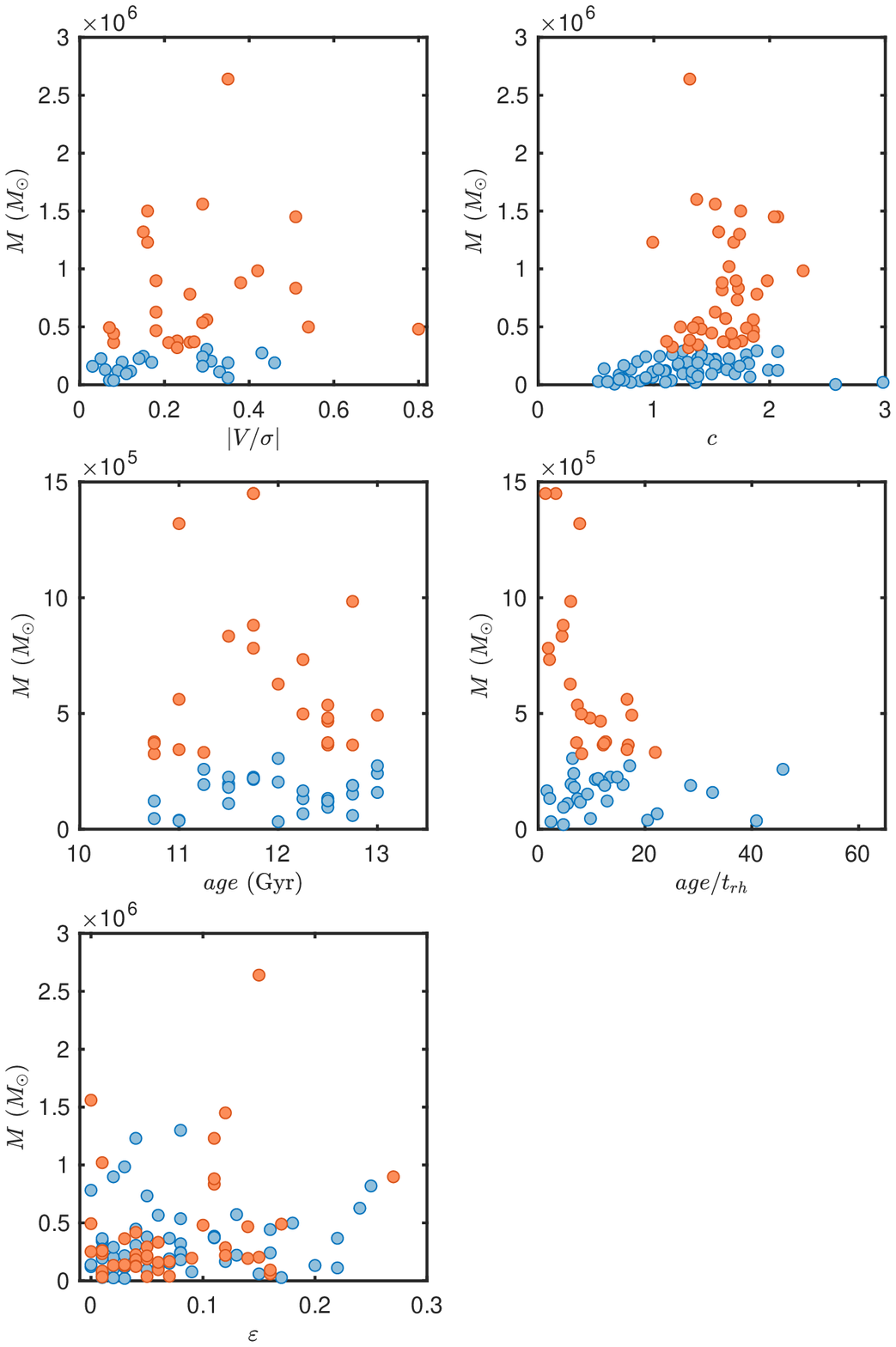}
\caption{The GC masses are plotted against the clusters $|V/\sigma|$ parameter (top left panel), the concentration (top right panel), the age (middle left panel), the dynamical age (middle right panel) and the ellipticity (bottom left panel) of the clusters.
Orange are clusters with masses larger than $10^{5.5}~M_\odot$ and light blue are clusters with masses smaller than the same threshold. Errors for the masses are not available for the data-set used.}\label{fig:mass}
\end{figure*}
%%%%%%%%%%%%%%%%%%%%%%%%%%%%%%%
%\begin{figure}\centering
%\includegraphics[width=0.4\textwidth]{rh_rtperi_F2}\\
%\includegraphics[width=0.415\textwidth]{rtperi_F2}\\
%\includegraphics[width=0.42\textwidth]{age_Tr_F2.eps}\quad
%\caption{The fraction of 2P stars, $F_2$, is plotted against 
%%the ratio between the half-mass radius and the tidal radius at the pericentre ($r_h/r_{t,p}$, top %anel), 
%the tidal radius at the pericentre (top panel) and the ratio between the age and the period of the %clusters (bottom panel). }\label{fig:rt_peri}
%\end{figure} 
%%%%%%%%%%%%%%%%%%%%%%%%%%%%%%%

\subsection{Multiple populations and their major correlations with cluster properties}
\label{sec:majorcorr}

Figure \ref{fig:strongcorr} shows the fraction of second generation stars, $F_2$ \citep[quantity available for $\sim 1/3$ of the Galactic GCs][]{MI17},  as a function of different Galactic GC internal parameters. In this figure, we present the strongest correlations found, while the full information on all the statistical parameters can be found in Table \ref{tab:tab1}.

{\bf GC masses}: As already found by \cite{MI17} and \cite{Mi19}, we confirm that $F_2$ increases with the total mass of the clusters (top left panel of Figure \ref{fig:strongcorr}).  In our analysis we adopted the masses given by \cite{GN97}. However, we obtain similar results when using the more recent mass compilation provided by \cite{BH18}.
As shown by the coefficients listed in  Table \ref{tab:tab1}, this correlation is tight (the correlation coefficient ranges between 0.58 and 0.77, depending on the statistical test being performed). In self-enrichment scenarios, more massive clusters can provide and retain more material  for an additional star formation event, leading naturally to a positive correlation between the mass and $F_2$.
 For masses smaller than $10^{5.5}~M_\odot$ the relationship shows a significant scatter and becomes bimodal. The  four main outliers (NGC 6366, NGC 2298, NGC 4833,  NGC 6681) show a larger $F_2$ compared to other clusters of similar mass. Those clusters are significantly dynamically evolved clusters; in particular, NGC 6366 and NGC 2298 are heavily stripped clusters \citep{Pa09, DM07}, the orbit of NGC 4833 suggests that bulge shocking removed a large fraction of the initial mass of this cluster \citep{Ca14} and NGC 6681 is a core collapsed cluster \citep{H96}. Moreover, the peri-Galactic distance of these clusters  is relatively small \citep[][]{Helmi18}, confirming that they could have lost a significant fraction of their original stellar population residing in the cluster outskirts due to the tidal interaction with the Galaxy \citep[see also][for a study of the correlation between $F_2$ and the peri-galactic distance of the clusters]{Mi19}.  If 2P stars form in the cluster's core, GCs that lost a large fraction of their initial mass are expected to have  larger $F_2$. In this scenario, the less concentrated 1P was, indeed, more easily stripped from the outskirts of the cluster because of evaporation and tidal interaction with the Galactic potential. \\
 
{\bf Concentration parameter:} $F_2$ strongly correlates with the concentration parameter --  an indicator of the depth of the central potential of the cluster (see top right panel of Figure \ref{fig:strongcorr}).  The correlation exists for the whole sample of GCs as well as for the clusters less massive than $10^{5.5}~M_\odot$.
If 2P stars form in the cluster core this population will be more bound in more concentrated clusters, and therefore less affected by tidal effects than 1P stars. 
 While low mass clusters cover a large range of $F_2$, massive clusters, which can retain a large fraction of gas, have 2P fractions that only vary between 0.6 and 0.8.
The two main outliers which show higher concentration than expected, NGC 6717 and NGC 7078, are both core collapsed clusters. Therefore, their concentration parameters are not accurate and should be considered only as indicative \citep{H96}. The main outlier showing a concentration smaller than expected is NGC 5139 ($\omega$ Cen), which is strongly suspected to be the former nucleus of an accreted and destroyed dwarf galaxy \citep{Nor95}.  \\

{\bf GC rotations}: While \cite{BP18} used Gaia data to provide the rotation signal of clusters on the plane of the sky ($v_t$), \cite{SBH19} combined Gaia proper motions to the most comprehensive set of line-of-sight velocities of Galactic GCs to obtain their 3D rotational amplitudes ($v_A$).
We plotted $F_2$ against both these quantities, finding a positive correlation (see bottom panels of Figure \ref{fig:strongcorr}). The large scatter and bimodality observed when using $v_t$ is reduced when plotting $F_2$ against $v_A$. The non rotating clusters, i.e. the ones showing rotational signal close to zero, show in both cases a large scatter. The scatter is reduced for clusters with higher rotational velocities. A relationship between $F_2$ and the rotational velocity is expected if 2P stars form from gas lost by 1P stars in a rotating cluster \citep[see e.g.][]{HB15,MBP16}. \\

 By considering a linear combination of the logarithm of the cluster masses ($M_i$), concentrations ($c_i$) amd rotational velocities ($v_{A,i}$)  
\begin{equation}
F_2=\alpha_0+\alpha_1 \log(M_i)+\alpha_2 c_i +\alpha_3 v_{A,i},
\end{equation}
we performed a linear regression analysis that identified the mass as the most significant parameter determining $F_2$ ($\alpha_1=5.6$, $p_{\rm value}=2.1\times10^{-6}$). The concentration is more significant  ($\alpha31=1.3$, $p_{\rm{value}}=0.20$) than the rotational velocity ($\alpha_3=0.11$, $p_{\rm value}=0.91$).  As only the $p_{\rm{value}}$ for the mass is smaller than $0.05$ and significantly smaller than the probabilities found for the other parameters, it appears that the cluster mass determines the 2P fraction, with little, if any, independent contribution from the other parameters, i.e. the observed correlations with the other parameters could arise solely from their correlation with the cluster mass.

 All the  correlations discussed above have high statistical significance at least for the entire sample of GCs and are suggestive of the fact that the 2P forms centrally concentrated inside the 1P cluster and the fraction of 2P stars is mainly correlated to the mass. 
There is also a correlation with the amplitude of tidal stripping experienced by the clusters  during their evolution, through their concentration parameter. 
The stronger the tidal effects, the more 1P stars -- which are less concentrated -- are stripped away increasing the relative 2P fraction, mostly distributed within the central region of  the cluster, in the remnant stellar population.  In such a case, the relaxation time has only a small influence on this process, at least for the less massive clusters (see Sect. \ref{sect:otherpars}). For shorter relaxation times we might expect a stronger mixing between the populations and therefore a somewhat lower increase of $F_2$ due to stripping (see Sect \ref{sect:otherpars}).\\

\subsection{Multiple populations minor correlations with cluster properties \label{sect:otherpars}}
Figure \ref{fig:other_corr} shows $F_2$ plotted against additional cluster parameters.  The statistical significance of the possible correlations existing between these quantities is listed in Table \ref{tab:tab2}.\\
The $F_2$ vs $|V/\sigma|$\footnote{The  $|V/\sigma|$ parameter is a measure of the ordered motion with respect to the velocity dispersion. As such, it gives an estimate of the rotation strength independently of the cluster mass.} plot shows several sequences and a positive trend. However, our tests do not detect a statistically significant correlation between these quantities, except for a slight trend which is, however, found only when considering the entire GC sample.  We note that \cite{Bellazzini12} found a slight trend between the extent of the Na-O anticorrelation  and $V/\sigma$. \\
Although by eye there seem to be a positive trend of $F_2$ with the ellipticity of the clusters ($\varepsilon$),  this qualitative observation is statistically confirmed only for the high mass clusters. 
The lack of correlation for the entire sample of GCs and for low mass clusters might be due to the large spread F2 among low mass GCs with similar ellipticities and to the existence of several outliers.\\
 While he age of the clusters is not correlated to the 2P fraction, $F_2$ shows an anticorrelation with the dynamical age of the clusters (i.e. the age of the cluster in terms of half-mass relaxation times, ${age}/t_{\rm rh}$) when considering the entire sample of clusters and only in the case of the Pearson statistical test.  In Appendix \ref{app:age} we discuss the implication of using the different age compilations available in literature \citep{FO10,Gratton10,Do10}.\\
$F_2$ and the half-mass relaxation time ($t_{\rm rh}$)  of the high mass clusters are correlated. The detected correlation is stronger when performing a Pearson and Spearman statistical test.\\
We also checked for the existence of a possible correlation between the number of orbits, i.e. of peri-centre passages, travelled by cluster and $F_2$. The number of orbits has been evaluated as the ratio between the age and the radial period, $T_r$, of the cluster taken from Model 1 in \cite{Helmi18}. 
 Interestingly, these two quantities are anticorrelated when taking into account  the lower mass clusters. The same anticorrelation is found when using the ages provided by \cite{Gratton10} but not when using the ages from \cite{Do10} and \cite{FO10}.

\subsection{Correlations between cluster rotations, ages and masses}\label{sec:rotcorr}
We checked for the existence of mutual correlations between the internal cluster parameters focusing on the rotation, concentration, mass,  age related parameters and ellipticity (see Figures \ref{fig:vtcorr} and \ref{fig:mass} as well as Table \ref{tab:tab2}).\\
The 3D rotational amplitude ($v_A$) is strongly correlated with the cluster mass. The concentration and $v_A$ are correlated only when considering the whole GC sample and in the case of the Kendal statistical test performed for low mass clusters.\\
The 3D rotational amplitude is not correlated  with the age of the clusters. Interestingly, $v_A$, is  anti-correlated with the dynamical age of the clusters when considering the entire sample of GCs or high mass clusters only.  Since clusters with larger relaxation times have typically smaller dynamical ages, our result is in agreement with the positive correlation found between the rotational velocities and the cluster relaxation times by \cite{Kamann18}. The amplitude of the rotational velocity is correlated with the ellipticity of the clusters ($\varepsilon$) for the entire GC sample only when removing the outliers. We should, however, caution that here we are considering the  projected ellipticities. Therefore, this result could be significantly affected by  projection effects that may smear underlying correlations. 
\\
The $V/\sigma$ parameter, which measures the strength of the rotation of the system compared to its velocity dispersion, seems to be mildly correlated with the current cluster mass as also found by \cite{BP18} and \cite{SBH19}. This is true only for the whole sample of GCs and after removing the effect of the outliers. 
The concentration and the mass are strongly correlated  when considering the whole sample of GCs as well as low mass GCs only. \\
 The mass and the age are not correlated.
However, the mass is strongly correlated with the dynamical age of high mass clusters.
Finally, the mass is not significantly correlated with the ellipticity of the clusters.

%%%%%%%%%%%%%%%%%%%%%%%%
\begin{table*}\centering
\small
\begin{tabular}{ccccccccc}
\hline
Par$_{1}$ & Par$_{2}$ & Stat & PCC & PP & PCC$_{LV}$ & PP$_{LV}$ &PCC$_{HV}$ & PP$_{HV}$\tabularnewline
\hline 

$F_2$ & $\log(M/r_c)$ & P &0.47& $1.5\times10^{-4}$& 0.41 & $0.0019$ &0.13& $0.55$ \tabularnewline
 & & S &0.67 & $2.9\times10^{-8}$& $0.49$& $0.0074$ & $0.24$ & 0.24 \tabularnewline
 & & K &$0.48$& $3.1\times10^{-7}$& $0.26$& $0.0062$ & $0.15$ & 0.30 \tabularnewline [0.1cm]

$F_2$ & $\log(M/r_c)$ & P &0.45 & $0.0027$ & 0.43 & $0.034$ & 0.11& $0.67$\tabularnewline
 &    & S & $0.71$ & $1.2\times10^{-7}$ & $0.56$ & 0.0039 & $-0.0098$ &  0.97 \tabularnewline
 &    & K & $0.53$ & $1.1\times10^{-6}$ & $0.41$ & 0.0044 & $-0.059$ &  0.78 \tabularnewline [0.1cm]

\hline 
\end{tabular}
\caption{Statistical correlation coefficients (PCC) and probabilities (PP) between $F_2$ and an escape velocity indicator calculated as the mass of the cluster over its core radius ($M/r_c$). Parameters are reported for the entire sample of GCs, for clusters with $M/r_c$  smaller (LV) and higher (HV) than $10^{6}~M_\odot/{\rm pc}$. P stands for Pearson, S for Spearman and K for Kendall.  The first three rows are for the entire sample of GCs, the last three are calculated excluding core collapsed clusters and clusters suspected to be accreted NSCs.}
\label{tab:tab3}
\end{table*}

\begin{figure}\centering
\includegraphics[width=0.45\textwidth]{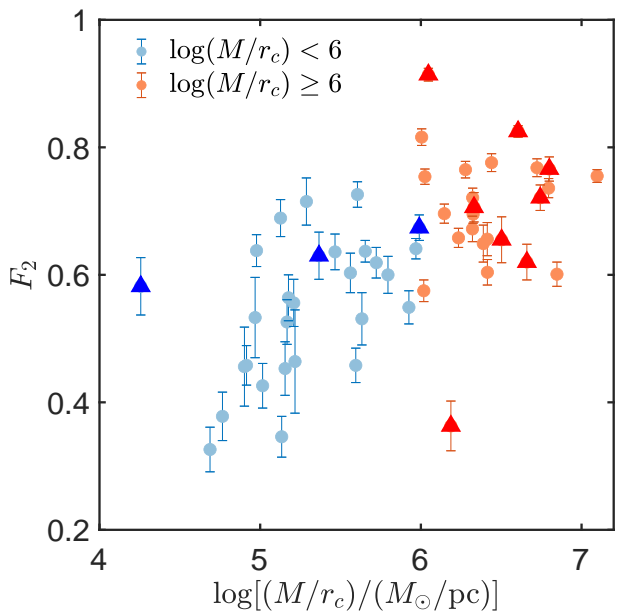}
\caption{The second generation fraction, $F_2$ is plotted against an indicator of the escape velocity of the cluster, $\log(M/r_c)$. Light blue are clusters with $\log(M/r_c)<6$, orange are clusters with $\log(M/r_c)\geq 6$. Triangles are core collapsed clusters and clusters suspected to be accreted NSCs. }\label{fig:escape_vel}
\end{figure}
%%%%%%%%%%%%%%%%%%%%%%%

\section{Discussion and Conclusions}\label{sec:con}
 GCs with ages larger than their relaxation times should still show signatures of their primordial structure, including those left by the 2P stars formation mechanism \citep[see e.g.][]{Vesperini13,MBP13,HB15,MBP16, CHB17, TV19, Vesperini21}.

By studying the relationships between
global parameters  (e.g. the HB morphology and structural or orbital parameters) and the abundance patterns of the different populations within each GC,
\cite{Car10} revised the GC formation mechanism dividing it into three phases; the formation
of a precursor population is followed by the birth of the primordial population and then
by the formation of the current GC, mainly from gas lost through the slow winds of a fraction of the primordial population.
This scenario implies that the initial clusters were several times more massive than the current ones and that the halo of the Galaxy could 
contain a large fraction of former GC stars. This study relied on a set of homogeneous chemical abundances
obtained for more than 1200 red giants in 19 clusters, as well as on additional data from literature and parameters coming from \cite[][2010 edition]{H96}.
 Through their approach, \cite{Car10}, found that the extent of the Na-O anticorrelation, calculated as the interquartile range (IQR[O/Na]), is mainly linked to the mass, age and metallicity of the clusters. \cite{Car19} used 22 GCs  observed with FLAMES to calibrate the observed IQR[O/Na] values with respect to structural parameters and HB morphology of 120 Galactic GCs. In such a way, they found that the main parameters linked to the 2P phenomenon are the initial mass and the concentration of the clusters.
 \cite{Bellazzini12} found a hint of a negative trend between IQR[O/Na]  and $V/\sigma$. However, they did not detect any correlation between the 2P and the rotational velocities of the observed clusters.\\
 \cite{MI17} and later \cite{Mi19} found a strong anti-correlation between the 1P stellar fraction and both the mass of the 2P component and the cluster initial mass. They found as well a milder correlation between the current cluster mass and that of the 2P \citep[see also][]{Da20}. The luminosity and peri-galactic radii of the clusters also  might play a role being correlated with the 1P fraction and helium variations respectively.
\cite{Da20} recently found that two clusters in the MCs show a smaller fraction of 2P stars, suggesting a link between the multiple population phenomenon and  environmental effects as the strength of the tidal forces exerted by host galaxy on the clusters.\\

In this work, we collected GC internal parameters from literature to search for possible correlations with the fraction of second generation stars observed in each cluster.  For the first time, we carry out an homogeneous statistical search to study all of these various parameters together and on the same extended sample.\\
We confirm the existence of a strong correlation between the second generation number fraction and the cluster mass. This tight relationship becomes bi-modal and more scattered at lower cluster masses.  This clear effect could be linked to the cluster conditions at the epoch of 2P formation and to their subsequent evolution.
A positive trend of $F_2$ with the present-day and initial mass of the clusters has been already observed  by \cite{MI17}, \cite{LG19}, \cite{Car19} and \cite{Mi19}, \cite{Gratton19} and it is expected if 2P stars form from material lost from 1P stars.
\\
We observe a  correlation between the second population stellar fraction and the rotational velocity of the observed clusters. Such correlation is expected if 2P stars form centrally concentrated from gas lost by 1P stars collected in a rotating disc at the cluster centre\footnote{ Qualitatively, we expect that when a larger quantity of gas (bringing in the orbital angular momentum inherited from the donor) collects at the centre of the cluster, a larger net angular momentum should be gained by the forming 2P stars. A growing F2 should, therefore, correspond to a growing rotational velocity. } \citep{HB15,MBP16}. To verify a direct connection with the formation process it is necessary to check the rotation of the 1P and 2P separately, to identify a differential rotation between the two populations. This has been done for few clusters, however, since the analysed clusters show significantly different behaviours, no firm conclusion has been yet achieved  \citep{Mi18, Co20b, Co20, Szigeti21}.
Additionally,  the strong correlation existing between cluster mass and rotation makes it difficult to disentangle between the relevance of these two properties in respect to the 2P origin. 
\\
A larger $F_2$ is observed for clusters with growing concentrations. In a scenario where 2P form at the cluster centre, a larger concentration corresponds to a more centrally bound second stellar population and to a stronger tidal stripping affecting the more radially extended 1P component. Therefore, a more significant 1P mass-loss  is expected compared with less concentrated clusters. 
 As previously also pointed out by \cite{RH13}, \cite{BV15}, \cite{CHB17}, \cite{Kamann20} and \cite{Da19}, the correlation existing between $F_2$ and the concentration parameter  supports the idea that 2P stars are born concentrated at the centre of the cluster, while the initial radial distribution of 1P stars is more extended. The more the cluster is concentrated the more 1P stars will be stripped away.   This possibility has been mostly dynamically and hydro-dynamically tested  in the case of the AGB scenario \citep[see e.g.][]{Dercole08, Bekki10, Bekki11, HB15,MBP13,MBP16,TV19, Calura19, Vesperini21}/ However, such relationship might be expected in any self-enrichment scenario.\\
 Our regression analysis  points at the mass as the main parameter that determines the fraction of second population stars. Other correlations observed with $F_2$  appear to be mediated by the mass e.g. with the rotational velocity or the concentration. 
\\

  While \cite{LG19} find a marginal correlation between the presence of second population stars and ages of Galactic and Magellanic clouds clusters, \cite{Martocchia18} and \cite{Ma19} results suggest that the age has a crucial role in shaping the characteristics of multiple populations. Older clusters, in fact, show larger nitrogen spreads compared to younger ones. However, with the available data we do not detect any correlation between the cluster age, and $F_2$.  \\
In principle, clusters which experience a larger number of peri-centre passages might be somewhat more stripped, though such an effect could be secondary compared with other parameters.  The number of peri-centre passages is measured by the age of the clusters divided by their radial orbital periods around the Galaxy, $T_r$. 
 We find that low mass clusters show a correlation between $F_2$ and $age/T_r$. In general, low mass clusters are more affected by tidal stripping than high mass clusters possibly explaining why we observe the correlation only for clusters less massive than $10^{5.5}~M_\odot$.
The effect of the tidal stripping on the fraction of second population stars is also confirmed by the fact that clusters with large perigalactic distances host, on average, a smaller $F_2$ \citep{Ze19}.  In contrast, \cite{Dercole08} and, more recently, \cite{Vesperini21} using $N$-body simulations of the cluster evolution, suggest that stellar evolution and primordial mass segregation could be the major causes of the preferential loss of 1P stars.\\ 

$F_2$ is mildly anti-correlated with the dynamical age (measured in relaxation times), especially  when applying a Pearson statistical test to the entire sample of clusters. This can be understood if 1P and 2P stars in dynamically older clusters are more mixed than in younger clusters.  In dynamically older GCs, then, tidal stripping can become more efficient in removing 2P stars which, in such clusters, are more abundant in the cluster outskirts. \\
 We also verified the existence of a relation between the $A^+$ parameter provided by \cite{Da19}, which is a measure of the 2P concentration with respect to the 1P, and $F_2$. We found that these two quantities are indeed anticorrelated, meaning that more concentrated 2P (i.e. less mixed populations or dynamically younger clusters) show lower $F_2$. 
 $F_2$ and the half-mass relaxation time ($t_{\rm rh}$) are correlated for high mass clusters. For these clusters, relaxation effects might have been more important than the tidal interaction with the Galaxy.
\\
 All the detected (anti)correlations appear to be bimodal, with different behaviours for clusters less massive or more massive than $10^{5.5}~M_\odot$. These include the mass correlation, which is stronger for more massive clusters. The mass cut-off could be linked to the formation mechanism, which seems to be more efficient for more massive clusters, and also to the following evolution which affects differently clusters roughly below and above the mass cut-off. Metal poor clusters ($[Fe/H]<-1$), i.e. possibly GCs accreted during mergers, show slightly stronger correlations with the mass, concentration and rotational velocity than  metal rich clusters ($[Fe/H]\geq -1$). The remaining correlations do not show any trend with the metallicity (see Appendix \ref{sec:app1}). \\

As a final test, we checked for the existence of a possible link between $F_2$ and the escape velocity of the clusters, estimated through an indicator ($M/r_c$, where $M$ is the mass of the cluster, and $r_c$ it is its core radius). \\
As shown in Figure \ref{fig:escape_vel} and Table \ref{tab:tab3}, all the clusters taken as a single group show a strong correlation between the 2P fraction and the escape velocity of the clusters.  A similar relationship has been also pointed out by \cite{BH18}.
 Additionally, we observe that while clusters with $M/r_c\geq 10^6~M_\odot/{\rm pc}$ do not show any correlation between these quantities, clusters with low escape velocity ($M/r_c<10^6~M_\odot/{\rm pc}$) display a correlation between $F_2$ and $\log(M/r_c)$.
This correlation that, for the high escape velocity clusters, is as strong as the one observed between $F_2$ and the cluster mass becomes 
slightly stronger when excluding core collapsed clusters \citep[][]{H96} and GCs suspected to be NSCs \citep{Pf20}. 

In self-enrichment scenarios, the fraction of retained gas and, therefore $F_2$, should increase with the escape velocity from the cluster, i.e. with the depth of the central potential well. However, sufficiently massive and centrally concentrated clusters have escape velocities high enough as to potentially retain all the gas lost through stellar winds from evolved stars, leading to second population fractions showing only small fluctuations around the mean value. 

 In conclusion, higher escape velocities, as well as higher masses, seem to correspond to a larger fraction of retained gas which, in turn, gives rise to larger fractions of 2P stars. The gas retained in the cluster core forms 2P stars, while 1P are more extended and more easily lost by the cluster in the interaction with the Galactic tidal field.

\section*{Acknowledgements}
The authors thank the referee for the helpful and careful comments that improved our work and A. F. Marino and A. Milone for the useful discussion and comments.
AMB is supported by the Swedish Research Council (grant 2017-04217).

\section*{Data availability}
The data underlying this article are available in the article and in the cited literature.
%%%%%%%%%%%%%%%%%%%%%%%%%%%%%%%%%%%%%%%%%%

\bibliographystyle{mnras}
\bibliography{gcm} % if your bibtex file is called example.bib

\begin{thebibliography}{}
\makeatletter
\relax
\def\mn@urlcharsother{\let\do\@makeother \do\$\do\&\do\#\do\^\do\_\do\%\do\~}
\def\mn@doi{\begingroup\mn@urlcharsother \@ifnextchar [ {\mn@doi@}
  {\mn@doi@[]}}
\def\mn@doi@[#1]#2{\def\@tempa{#1}\ifx\@tempa\@empty \href
  {http://dx.doi.org/#2} {doi:#2}\else \href {http://dx.doi.org/#2} {#1}\fi
  \endgroup}
\def\mn@eprint#1#2{\mn@eprint@#1:#2::\@nil}
\def\mn@eprint@arXiv#1{\href {http://arxiv.org/abs/#1} {{\tt arXiv:#1}}}
\def\mn@eprint@dblp#1{\href {http://dblp.uni-trier.de/rec/bibtex/#1.xml}
  {dblp:#1}}
\def\mn@eprint@#1:#2:#3:#4\@nil{\def\@tempa {#1}\def\@tempb {#2}\def\@tempc
  {#3}\ifx \@tempc \@empty \let \@tempc \@tempb \let \@tempb \@tempa \fi \ifx
  \@tempb \@empty \def\@tempb {arXiv}\fi \@ifundefined
  {mn@eprint@\@tempb}{\@tempb:\@tempc}{\expandafter \expandafter \csname
  mn@eprint@\@tempb\endcsname \expandafter{\@tempc}}}

\bibitem[\protect\citeauthoryear{{Bastian} \& {Lardo}}{{Bastian} \&
  {Lardo}}{2018}]{BL18}
{Bastian} N.,  {Lardo} C.,  2018, \mn@doi [\araa]
  {10.1146/annurev-astro-081817-051839}, \href
  {http://adsabs.harvard.edu/abs/2018ARA%26A..56...83B} {56, 83}

\bibitem[\protect\citeauthoryear{{Bastian}, {Lamers}, {de Mink}, {Longmore},
  {Goodwin}  \& {Gieles}}{{Bastian} et~al.}{2013}]{BA13}
{Bastian} N.,  {Lamers} H.~J.~G.~L.~M.,  {de Mink} S.~E.,  {Longmore} S.~N.,
  {Goodwin} S.~P.,   {Gieles} M.,  2013, \mn@doi [\mnras]
  {10.1093/mnras/stt1745}, \href
  {http://adsabs.harvard.edu/abs/2013MNRAS.436.2398B} {436, 2398}

\bibitem[\protect\citeauthoryear{{Baumgardt} \& {Hilker}}{{Baumgardt} \&
  {Hilker}}{2018}]{BH18}
{Baumgardt} H.,  {Hilker} M.,  2018, \mn@doi [\mnras] {10.1093/mnras/sty1057},
  \href {http://adsabs.harvard.edu/abs/2018MNRAS.478.1520B} {478, 1520}

\bibitem[\protect\citeauthoryear{{Bekki}}{{Bekki}}{2010}]{Bekki10}
{Bekki} K.,  2010, \mn@doi [\apjl] {10.1088/2041-8205/724/1/L99}, \href
  {https://ui.adsabs.harvard.edu/abs/2010ApJ...724L..99B} {724, L99}

\bibitem[\protect\citeauthoryear{{Bekki}}{{Bekki}}{2011}]{Bekki11}
{Bekki} K.,  2011, \mn@doi [\mnras] {10.1111/j.1365-2966.2010.18047.x}, \href
  {https://ui.adsabs.harvard.edu/abs/2011MNRAS.412.2241B} {412, 2241}

\bibitem[\protect\citeauthoryear{{Bellazzini}, {Bragaglia}, {Carretta},
  {Gratton}, {Lucatello}, {Catanzaro}  \& {Leone}}{{Bellazzini}
  et~al.}{2012}]{Bellazzini12}
{Bellazzini} M.,  {Bragaglia} A.,  {Carretta} E.,  {Gratton} R.~G.,
  {Lucatello} S.,  {Catanzaro} G.,   {Leone} F.,  2012, \mn@doi [\aap]
  {10.1051/0004-6361/201118056}, \href
  {https://ui.adsabs.harvard.edu/abs/2012A&A...538A..18B} {538, A18}

\bibitem[\protect\citeauthoryear{{Bellini} et~al.,}{{Bellini}
  et~al.}{2015}]{BV15}
{Bellini} A.,  et~al., 2015, \mn@doi [\apjl] {10.1088/2041-8205/810/1/L13},
  \href {https://ui.adsabs.harvard.edu/abs/2015ApJ...810L..13B} {810, L13}

\bibitem[\protect\citeauthoryear{{Bellini} et~al.,}{{Bellini}
  et~al.}{2018}]{BLB18}
{Bellini} A.,  et~al., 2018, \mn@doi [\apj] {10.3847/1538-4357/aaa3ec}, \href
  {https://ui.adsabs.harvard.edu/abs/2018ApJ...853...86B} {853, 86}

\bibitem[\protect\citeauthoryear{{Bianchini}, {van der Marel}, {del Pino},
  {Watkins}, {Bellini}, {Fardal}, {Libralato}  \& {Sills}}{{Bianchini}
  et~al.}{2018}]{BP18}
{Bianchini} P.,  {van der Marel} R.~P.,  {del Pino} A.,  {Watkins} L.~L.,
  {Bellini} A.,  {Fardal} M.~A.,  {Libralato} M.,   {Sills} A.,  2018, \mn@doi
  [\mnras] {10.1093/mnras/sty2365}, \href
  {http://adsabs.harvard.edu/abs/2018MNRAS.481.2125B} {481, 2125}

\bibitem[\protect\citeauthoryear{{Breen}}{{Breen}}{2018}]{BR18}
{Breen} P.~G.,  2018, \mn@doi [\mnras] {10.1093/mnrasl/sly169}, \href
  {http://adsabs.harvard.edu/abs/2018MNRAS.481L.110B} {481, L110}

\bibitem[\protect\citeauthoryear{{Calura}, {D'Ercole}, {Vesperini}, {Vanzella}
  \& {Sollima}}{{Calura} et~al.}{2019}]{Calura19}
{Calura} F.,  {D'Ercole} A.,  {Vesperini} E.,  {Vanzella} E.,   {Sollima} A.,
  2019, \mn@doi [\mnras] {10.1093/mnras/stz2055}, \href
  {https://ui.adsabs.harvard.edu/abs/2019MNRAS.489.3269C} {489, 3269}

\bibitem[\protect\citeauthoryear{{Carretta}}{{Carretta}}{2019}]{Car19}
{Carretta} E.,  2019, \mn@doi [\aap] {10.1051/0004-6361/201935110}, \href
  {https://ui.adsabs.harvard.edu/abs/2019A&A...624A..24C} {624, A24}

\bibitem[\protect\citeauthoryear{{Carretta} et~al.,}{{Carretta}
  et~al.}{2009}]{CBG09}
{Carretta} E.,  et~al., 2009, \mn@doi [\aap] {10.1051/0004-6361/200912096},
  \href {https://ui.adsabs.harvard.edu/abs/2009A&A...505..117C} {505, 117}

\bibitem[\protect\citeauthoryear{{Carretta}, {Bragaglia}, {Gratton},
  {Recio-Blanco}, {Lucatello}, {D'Orazi}  \& {Cassisi}}{{Carretta}
  et~al.}{2010}]{Car10}
{Carretta} E.,  {Bragaglia} A.,  {Gratton} R.~G.,  {Recio-Blanco} A.,
  {Lucatello} S.,  {D'Orazi} V.,   {Cassisi} S.,  2010, \mn@doi [\aap]
  {10.1051/0004-6361/200913451}, \href
  {http://adsabs.harvard.edu/abs/2010A\%26A...516A..55C} {516, A55}

\bibitem[\protect\citeauthoryear{{Carretta} et~al.,}{{Carretta}
  et~al.}{2014}]{Ca14}
{Carretta} E.,  et~al., 2014, \mn@doi [\aap] {10.1051/0004-6361/201323321},
  \href {https://ui.adsabs.harvard.edu/abs/2014A&A...564A..60C} {564, A60}

\bibitem[\protect\citeauthoryear{{Cordero}, {Pilachowski}, {Johnson},
  {McDonald}, {Zijlstra}  \& {Simmerer}}{{Cordero} et~al.}{2014}]{2CPJ14}
{Cordero} M.~J.,  {Pilachowski} C.~A.,  {Johnson} C.~I.,  {McDonald} I.,
  {Zijlstra} A.~A.,   {Simmerer} J.,  2014, \mn@doi [\apj]
  {10.1088/0004-637X/780/1/94}, \href
  {https://ui.adsabs.harvard.edu/abs/2014ApJ...780...94C} {780, 94}

\bibitem[\protect\citeauthoryear{{Cordero}, {H{\'e}nault-Brunet},
  {Pilachowski}, {Balbinot}, {Johnson}  \& {Varri}}{{Cordero}
  et~al.}{2017}]{CHB17}
{Cordero} M.~J.,  {H{\'e}nault-Brunet} V.,  {Pilachowski} C.~A.,  {Balbinot}
  E.,  {Johnson} C.~I.,   {Varri} A.~L.,  2017, \mn@doi [\mnras]
  {10.1093/mnras/stw2812}, \href
  {https://ui.adsabs.harvard.edu/abs/2017MNRAS.465.3515C} {465, 3515}

\bibitem[\protect\citeauthoryear{{Cordoni}, {Milone}, {Mastrobuono-Battisti},
  {Marino}, {Lagioia}  \& {Tailo}}{{Cordoni} et~al.}{2019}]{CM19}
{Cordoni} G.,  {Milone} A.~P.,  {Mastrobuono-Battisti} A.,  {Marino} A.~F.,
  {Lagioia} E.~P.,   {Tailo} M.,  2019, arXiv e-prints, \href
  {https://ui.adsabs.harvard.edu/abs/2019arXiv190811692C} {p. arXiv:1908.11692}

\bibitem[\protect\citeauthoryear{{Cordoni}, {Milone}, {Mastrobuono-Battisti},
  {Marino}, {Lagioia}, {Tailo}, {Baumgardt}  \& {Hilker}}{{Cordoni}
  et~al.}{2020a}]{Co20b}
{Cordoni} G.,  {Milone} A.~P.,  {Mastrobuono-Battisti} A.,  {Marino} A.~F.,
  {Lagioia} E.~P.,  {Tailo} M.,  {Baumgardt} H.,   {Hilker} M.,  2020a, \mn@doi
  [\apj] {10.3847/1538-4357/ab5aee}, \href
  {https://ui.adsabs.harvard.edu/abs/2020ApJ...889...18C} {889, 18}

\bibitem[\protect\citeauthoryear{{Cordoni} et~al.,}{{Cordoni}
  et~al.}{2020b}]{Co20}
{Cordoni} G.,  et~al., 2020b, \mn@doi [\apj] {10.3847/1538-4357/aba04b}, \href
  {https://ui.adsabs.harvard.edu/abs/2020ApJ...898..147C} {898, 147}

\bibitem[\protect\citeauthoryear{{D'Ercole}, {Vesperini}, {D'Antona},
  {McMillan}  \& {Recchi}}{{D'Ercole} et~al.}{2008}]{Dercole08}
{D'Ercole} A.,  {Vesperini} E.,  {D'Antona} F.,  {McMillan} S. L.~W.,
  {Recchi} S.,  2008, \mn@doi [\mnras] {10.1111/j.1365-2966.2008.13915.x},
  \href {https://ui.adsabs.harvard.edu/abs/2008MNRAS.391..825D} {391, 825}

\bibitem[\protect\citeauthoryear{{Dalessandro} et~al.,}{{Dalessandro}
  et~al.}{2018}]{DM18}
{Dalessandro} E.,  et~al., 2018, \mn@doi [\apj] {10.3847/1538-4357/aad4b3},
  \href {https://ui.adsabs.harvard.edu/abs/2018ApJ...864...33D} {864, 33}

\bibitem[\protect\citeauthoryear{{Dalessandro} et~al.,}{{Dalessandro}
  et~al.}{2019}]{Da19}
{Dalessandro} E.,  et~al., 2019, \mn@doi [\apjl] {10.3847/2041-8213/ab45f7},
  \href {https://ui.adsabs.harvard.edu/abs/2019ApJ...884L..24D} {884, L24}

\bibitem[\protect\citeauthoryear{{Decressin}, {Meynet}, {Charbonnel},
  {Prantzos}  \& {Ekstr{\"o}m}}{{Decressin} et~al.}{2007}]{De07}
{Decressin} T.,  {Meynet} G.,  {Charbonnel} C.,  {Prantzos} N.,   {Ekstr{\"o}m}
  S.,  2007, \mn@doi [\aap] {10.1051/0004-6361:20066013}, \href
  {http://adsabs.harvard.edu/abs/2007A\%26A...464.1029D} {464, 1029}

\bibitem[\protect\citeauthoryear{{Dondoglio}, {Milone}, {Cordoni}, {Jang}  \&
  {Carlos}}{{Dondoglio} et~al.}{2020}]{Da20}
{Dondoglio} E.,  {Milone} A.~P.,  {Cordoni} G.,  {Jang} S.,   {Carlos} M.~G.,
  2020, arXiv e-prints, \href
  {https://ui.adsabs.harvard.edu/abs/2020arXiv201103283D} {p. arXiv:2011.03283}

\bibitem[\protect\citeauthoryear{{Dotter} et~al.,}{{Dotter}
  et~al.}{2010}]{Do10}
{Dotter} A.,  et~al., 2010, \mn@doi [\apj] {10.1088/0004-637X/708/1/698}, \href
  {https://ui.adsabs.harvard.edu/abs/2010ApJ...708..698D} {708, 698}

\bibitem[\protect\citeauthoryear{{Forbes} \& {Bridges}}{{Forbes} \&
  {Bridges}}{2010}]{FO10}
{Forbes} D.~A.,  {Bridges} T.,  2010, \mn@doi [\mnras]
  {10.1111/j.1365-2966.2010.16373.x}, \href
  {http://adsabs.harvard.edu/abs/2010MNRAS.404.1203F} {404, 1203}

\bibitem[\protect\citeauthoryear{{Gaia Collaboration} et~al.,}{{Gaia
  Collaboration} et~al.}{2018}]{Helmi18}
{Gaia Collaboration} et~al., 2018, \mn@doi [\aap]
  {10.1051/0004-6361/201832698}, \href
  {https://ui.adsabs.harvard.edu/abs/2018A&A...616A..12G} {616, A12}

\bibitem[\protect\citeauthoryear{{Gieles} et~al.,}{{Gieles}
  et~al.}{2018}]{GCK18}
{Gieles} M.,  et~al., 2018, \mn@doi [\mnras] {10.1093/mnras/sty1059}, \href
  {http://adsabs.harvard.edu/abs/2018MNRAS.478.2461G} {478, 2461}

\bibitem[\protect\citeauthoryear{{Gnedin} \& {Ostriker}}{{Gnedin} \&
  {Ostriker}}{1997}]{GN97}
{Gnedin} O.~Y.,  {Ostriker} J.~P.,  1997, \mn@doi [\apj] {10.1086/303441},
  \href {http://adsabs.harvard.edu/abs/1997ApJ...474..223G} {474, 223}

\bibitem[\protect\citeauthoryear{{Gratton}, {Carretta}, {Bragaglia},
  {Lucatello}  \& {D'Orazi}}{{Gratton} et~al.}{2010}]{Gratton10}
{Gratton} R.~G.,  {Carretta} E.,  {Bragaglia} A.,  {Lucatello} S.,   {D'Orazi}
  V.,  2010, \mn@doi [\aap] {10.1051/0004-6361/200912572}, \href
  {https://ui.adsabs.harvard.edu/abs/2010A&A...517A..81G} {517, A81}

\bibitem[\protect\citeauthoryear{{Gratton}, {Carretta}  \&
  {Bragaglia}}{{Gratton} et~al.}{2012}]{Gra12}
{Gratton} R.~G.,  {Carretta} E.,   {Bragaglia} A.,  2012, \mn@doi [A\&ARr]
  {10.1007/s00159-012-0050-3}, \href
  {http://adsabs.harvard.edu/abs/2012A\%26ARv..20...50G} {20, 50}

\bibitem[\protect\citeauthoryear{{Gratton}, {Bragaglia}, {Carretta}, {D'Orazi},
  {Lucatello}  \& {Sollima}}{{Gratton} et~al.}{2019}]{Gratton19}
{Gratton} R.,  {Bragaglia} A.,  {Carretta} E.,  {D'Orazi} V.,  {Lucatello} S.,
   {Sollima} A.,  2019, \mn@doi [\aapr] {10.1007/s00159-019-0119-3}, \href
  {https://ui.adsabs.harvard.edu/abs/2019A&ARv..27....8G} {27, 8}

\bibitem[\protect\citeauthoryear{{Harris}}{{Harris}}{1996}]{H96}
{Harris} W.~E.,  1996, \mn@doi [\aj] {10.1086/118116}, \href
  {http://adsabs.harvard.edu/abs/1996AJ....112.1487H} {112, 1487}

\bibitem[\protect\citeauthoryear{{H{\'e}nault-Brunet}, {Gieles}, {Agertz}  \&
  {Read}}{{H{\'e}nault-Brunet} et~al.}{2015}]{HB15}
{H{\'e}nault-Brunet} V.,  {Gieles} M.,  {Agertz} O.,   {Read} J.~I.,  2015,
  \mn@doi [\mnras] {10.1093/mnras/stv675}, \href
  {http://adsabs.harvard.edu/abs/2015MNRAS.450.1164H} {450, 1164}

\bibitem[\protect\citeauthoryear{{Kamann} et~al.,}{{Kamann}
  et~al.}{2018}]{Kamann18}
{Kamann} S.,  et~al., 2018, \mn@doi [\mnras] {10.1093/mnras/stx2719}, \href
  {https://ui.adsabs.harvard.edu/abs/2018MNRAS.473.5591K} {473, 5591}

\bibitem[\protect\citeauthoryear{{Kamann} et~al.,}{{Kamann}
  et~al.}{2020}]{Kamann20}
{Kamann} S.,  et~al., 2020, \mn@doi [\mnras] {10.1093/mnras/stz3506}, \href
  {https://ui.adsabs.harvard.edu/abs/2020MNRAS.492..966K} {492, 966}

\bibitem[\protect\citeauthoryear{{Lagioia}, {Milone}, {Marino}, {Cordoni}  \&
  {Tailo}}{{Lagioia} et~al.}{2019}]{LG19}
{Lagioia} E.~P.,  {Milone} A.~P.,  {Marino} A.~F.,  {Cordoni} G.,   {Tailo} M.,
   2019, \mn@doi [\aj] {10.3847/1538-3881/ab45f2}, \href
  {https://ui.adsabs.harvard.edu/abs/2019AJ....158..202L} {158, 202}

\bibitem[\protect\citeauthoryear{{Lee}}{{Lee}}{2017}]{LJ17}
{Lee} J.-W.,  2017, \mn@doi [\apj] {10.3847/1538-4357/aa7b8c}, \href
  {https://ui.adsabs.harvard.edu/abs/2017ApJ...844...77L} {844, 77}

\bibitem[\protect\citeauthoryear{{Lee}}{{Lee}}{2018}]{LJ18}
{Lee} J.-W.,  2018, \mn@doi [\apjs] {10.3847/1538-4365/aadcad}, \href
  {https://ui.adsabs.harvard.edu/abs/2018ApJS..238...24L} {238, 24}

\bibitem[\protect\citeauthoryear{{Martocchia} et~al.,}{{Martocchia}
  et~al.}{2018}]{Martocchia18}
{Martocchia} S.,  et~al., 2018, \mn@doi [\mnras] {10.1093/mnras/stx2556}, \href
  {https://ui.adsabs.harvard.edu/abs/2018MNRAS.473.2688M} {473, 2688}

\bibitem[\protect\citeauthoryear{{Martocchia} et~al.,}{{Martocchia}
  et~al.}{2019}]{Ma19}
{Martocchia} S.,  et~al., 2019, \mn@doi [\mnras] {10.1093/mnras/stz1596}, \href
  {https://ui.adsabs.harvard.edu/abs/2019MNRAS.487.5324M} {487, 5324}

\bibitem[\protect\citeauthoryear{{Mastrobuono-Battisti} \&
  {Perets}}{{Mastrobuono-Battisti} \& {Perets}}{2013}]{MBP13}
{Mastrobuono-Battisti} A.,  {Perets} H.~B.,  2013, \mn@doi [\apj]
  {10.1088/0004-637X/779/1/85}, \href
  {http://adsabs.harvard.edu/abs/2013ApJ...779...85M} {779, 85}

\bibitem[\protect\citeauthoryear{{Mastrobuono-Battisti} \&
  {Perets}}{{Mastrobuono-Battisti} \& {Perets}}{2016}]{MBP16}
{Mastrobuono-Battisti} A.,  {Perets} H.~B.,  2016, \mn@doi [\apj]
  {10.3847/0004-637X/823/1/61}, \href
  {http://adsabs.harvard.edu/abs/2016ApJ...823...61M} {823, 61}

\bibitem[\protect\citeauthoryear{{Milone} et~al.,}{{Milone}
  et~al.}{2012a}]{MPB12}
{Milone} A.~P.,  et~al., 2012a, \mn@doi [\aap] {10.1051/0004-6361/201016384},
  \href {https://ui.adsabs.harvard.edu/abs/2012A&A...540A..16M} {540, A16}

\bibitem[\protect\citeauthoryear{{Milone}, {Marino}, {Piotto}, {Bedin},
  {Anderson}, {Aparicio}, {Cassisi}  \& {Rich}}{{Milone} et~al.}{2012b}]{Mil12}
{Milone} A.~P.,  {Marino} A.~F.,  {Piotto} G.,  {Bedin} L.~R.,  {Anderson} J.,
  {Aparicio} A.,  {Cassisi} S.,   {Rich} R.~M.,  2012b, \mn@doi [\apj]
  {10.1088/0004-637X/745/1/27}, \href
  {http://adsabs.harvard.edu/abs/2012ApJ...745...27M} {745, 27}

\bibitem[\protect\citeauthoryear{{Milone} et~al.,}{{Milone}
  et~al.}{2016}]{Milone16}
{Milone} A.~P.,  et~al., 2016, \mn@doi [\mnras] {10.1093/mnras/stv2415}, \href
  {https://ui.adsabs.harvard.edu/abs/2016MNRAS.455.3009M} {455, 3009}

\bibitem[\protect\citeauthoryear{{Milone} et~al.,}{{Milone}
  et~al.}{2017}]{MI17}
{Milone} A.~P.,  et~al., 2017, \mn@doi [\mnras] {10.1093/mnras/stw2531}, \href
  {http://adsabs.harvard.edu/abs/2017MNRAS.464.3636M} {464, 3636}

\bibitem[\protect\citeauthoryear{{Milone}, {Marino}, {Mastrobuono-Battisti}  \&
  {Lagioia}}{{Milone} et~al.}{2018a}]{MM18}
{Milone} A.~P.,  {Marino} A.~F.,  {Mastrobuono-Battisti} A.,   {Lagioia} E.~P.,
   2018a, \mn@doi [\mnras] {10.1093/mnras/sty1873}, \href
  {https://ui.adsabs.harvard.edu/abs/2018MNRAS.479.5005M} {479, 5005}

\bibitem[\protect\citeauthoryear{{Milone}, {Marino}, {Mastrobuono-Battisti}  \&
  {Lagioia}}{{Milone} et~al.}{2018b}]{Mi18}
{Milone} A.~P.,  {Marino} A.~F.,  {Mastrobuono-Battisti} A.,   {Lagioia} E.~P.,
   2018b, \mn@doi [\mnras] {10.1093/mnras/sty1873}, \href
  {https://ui.adsabs.harvard.edu/abs/2018MNRAS.479.5005M} {479, 5005}

\bibitem[\protect\citeauthoryear{{Milone} et~al.,}{{Milone}
  et~al.}{2020}]{Mi19}
{Milone} A.~P.,  et~al., 2020, \mn@doi [\mnras] {10.1093/mnras/stz2999}, \href
  {https://ui.adsabs.harvard.edu/abs/2020MNRAS.491..515M} {491, 515}

\bibitem[\protect\citeauthoryear{{Norris} \& {Da Costa}}{{Norris} \& {Da
  Costa}}{1995}]{Nor95}
{Norris} J.~E.,  {Da Costa} G.~S.,  1995, \mn@doi [ApJ] {10.1086/175909}, \href
  {http://adsabs.harvard.edu/abs/1995ApJ...447..680N} {447, 680}

\bibitem[\protect\citeauthoryear{{Norris} \& {Freeman}}{{Norris} \&
  {Freeman}}{1979}]{NF79}
{Norris} J.,  {Freeman} K.~C.,  1979, \mn@doi [\apjl] {10.1086/182988}, \href
  {https://ui.adsabs.harvard.edu/abs/1979ApJ...230L.179N} {230, L179}

\bibitem[\protect\citeauthoryear{{Paust} et~al.,}{{Paust} et~al.}{2009}]{Pa09}
{Paust} N. E.~Q.,  et~al., 2009, \mn@doi [\aj] {10.1088/0004-6256/137/1/246},
  \href {https://ui.adsabs.harvard.edu/abs/2009AJ....137..246P} {137, 246}

\bibitem[\protect\citeauthoryear{{Pfeffer}, {Lardo}, {Bastian}, {Saracino}  \&
  {Kamann}}{{Pfeffer} et~al.}{2020}]{Pf20}
{Pfeffer} J.,  {Lardo} C.,  {Bastian} N.,  {Saracino} S.,   {Kamann} S.,  2020,
  arXiv e-prints, \href {https://ui.adsabs.harvard.edu/abs/2020arXiv201102042P}
  {p. arXiv:2011.02042}

\bibitem[\protect\citeauthoryear{{Piatti}}{{Piatti}}{2020}]{Piatti20}
{Piatti} A.~E.,  2020, \mn@doi [\aap] {10.1051/0004-6361/202039128}, \href
  {https://ui.adsabs.harvard.edu/abs/2020A&A...643A..77P} {643, A77}

\bibitem[\protect\citeauthoryear{{Piotto} et~al.,}{{Piotto}
  et~al.}{2007}]{PBA07}
{Piotto} G.,  et~al., 2007, \mn@doi [\apjl] {10.1086/518503}, \href
  {https://ui.adsabs.harvard.edu/abs/2007ApJ...661L..53P} {661, L53}

\bibitem[\protect\citeauthoryear{{Renzini} et~al.,}{{Renzini}
  et~al.}{2015}]{Ren15}
{Renzini} A.,  et~al., 2015, \mn@doi [\mnras] {10.1093/mnras/stv2268}, \href
  {http://adsabs.harvard.edu/abs/2015MNRAS.454.4197R} {454, 4197}

\bibitem[\protect\citeauthoryear{{Richer}, {Heyl}, {Anderson}, {Kalirai},
  {Shara}, {Dotter}, {Fahlman}  \& {Rich}}{{Richer} et~al.}{2013}]{RH13}
{Richer} H.~B.,  {Heyl} J.,  {Anderson} J.,  {Kalirai} J.~S.,  {Shara} M.~M.,
  {Dotter} A.,  {Fahlman} G.~G.,   {Rich} R.~M.,  2013, \mn@doi [\apjl]
  {10.1088/2041-8205/771/1/L15}, \href
  {https://ui.adsabs.harvard.edu/abs/2013ApJ...771L..15R} {771, L15}

\bibitem[\protect\citeauthoryear{{Sollima}, {Baumgardt}  \& {Hilker}}{{Sollima}
  et~al.}{2019}]{SBH19}
{Sollima} A.,  {Baumgardt} H.,   {Hilker} M.,  2019, \mn@doi [\mnras]
  {10.1093/mnras/stz505}, \href
  {http://adsabs.harvard.edu/abs/2019MNRAS.485.1460S} {485, 1460}

\bibitem[\protect\citeauthoryear{{Szigeti}, {M{\'e}sz{\'a}ros}, {Szab{\'o}},
  {Fern{\'a}ndez-Trincado}, {Lane}  \& {Cohen}}{{Szigeti}
  et~al.}{2021}]{Szigeti21}
{Szigeti} L.,  {M{\'e}sz{\'a}ros} S.,  {Szab{\'o}} G.~M.,
  {Fern{\'a}ndez-Trincado} J.~G.,  {Lane} R.~R.,   {Cohen} R.~E.,  2021,
  \mn@doi [\mnras] {10.1093/mnras/stab1007}, \href
  {https://ui.adsabs.harvard.edu/abs/2021MNRAS.tmp..982S} {}

\bibitem[\protect\citeauthoryear{{Tiongco}, {Vesperini}  \& {Varri}}{{Tiongco}
  et~al.}{2019}]{TV19}
{Tiongco} M.~A.,  {Vesperini} E.,   {Varri} A.~L.,  2019, \mn@doi [\mnras]
  {10.1093/mnras/stz1595}, \href
  {https://ui.adsabs.harvard.edu/abs/2019MNRAS.487.5535T} {487, 5535}

\bibitem[\protect\citeauthoryear{{VandenBerg}, {Brogaard}, {Leaman}  \&
  {Casagrand e}}{{VandenBerg} et~al.}{2013}]{VDB13}
{VandenBerg} D.~A.,  {Brogaard} K.,  {Leaman} R.,   {Casagrand e} L.,  2013,
  \mn@doi [\apj] {10.1088/0004-637X/775/2/134}, \href
  {https://ui.adsabs.harvard.edu/abs/2013ApJ...775..134V} {775, 134}

\bibitem[\protect\citeauthoryear{{Ventura}, {D'Antona}, {Mazzitelli}  \&
  {Gratton}}{{Ventura} et~al.}{2001}]{Ve01}
{Ventura} P.,  {D'Antona} F.,  {Mazzitelli} I.,   {Gratton} R.,  2001, \mn@doi
  [\apj] {10.1086/319496}, \href
  {https://ui.adsabs.harvard.edu/#abs/2001ApJ...550L..65V} {550, L65}

\bibitem[\protect\citeauthoryear{{Vesperini}, {McMillan}, {D'Antona}  \&
  {D'Ercole}}{{Vesperini} et~al.}{2013}]{Vesperini13}
{Vesperini} E.,  {McMillan} S. L.~W.,  {D'Antona} F.,   {D'Ercole} A.,  2013,
  \mn@doi [\mnras] {10.1093/mnras/sts434}, \href
  {https://ui.adsabs.harvard.edu/abs/2013MNRAS.429.1913V} {429, 1913}

\bibitem[\protect\citeauthoryear{{Vesperini}, {Hong}, {Giersz}  \&
  {Hypki}}{{Vesperini} et~al.}{2021}]{Vesperini21}
{Vesperini} E.,  {Hong} J.,  {Giersz} M.,   {Hypki} A.,  2021, \mn@doi [\mnras]
  {10.1093/mnras/stab223}, \href
  {https://ui.adsabs.harvard.edu/abs/2021MNRAS.502.4290V} {502, 4290}

\bibitem[\protect\citeauthoryear{{Wang}, {Kroupa}, {Takahashi}  \&
  {Jerabkova}}{{Wang} et~al.}{2020}]{Wang20}
{Wang} L.,  {Kroupa} P.,  {Takahashi} K.,   {Jerabkova} T.,  2020, \mn@doi
  [\mnras] {10.1093/mnras/stz3033}, \href
  {https://ui.adsabs.harvard.edu/abs/2020MNRAS.491..440W} {491, 440}

\bibitem[\protect\citeauthoryear{{Zennaro}, {Milone}, {Marino}, {Cordoni},
  {Lagioia}  \& {Tailo}}{{Zennaro} et~al.}{2019}]{Ze19}
{Zennaro} M.,  {Milone} A.~P.,  {Marino} A.~F.,  {Cordoni} G.,  {Lagioia}
  E.~P.,   {Tailo} M.,  2019, \mn@doi [\mnras] {10.1093/mnras/stz1477}, \href
  {https://ui.adsabs.harvard.edu/abs/2019MNRAS.487.3239Z} {487, 3239}

\bibitem[\protect\citeauthoryear{{de Marchi} \& {Pulone}}{{de Marchi} \&
  {Pulone}}{2007}]{DM07}
{de Marchi} G.,  {Pulone} L.,  2007, \mn@doi [\aap]
  {10.1051/0004-6361:20066719}, \href
  {https://ui.adsabs.harvard.edu/abs/2007A&A...467..107D} {467, 107}

\makeatother
\end{thebibliography}

\appendix

\section{Correlations and cluster metallicities} \label{sec:app1}
\begin{figure*}\centering
\includegraphics[width=0.85\textwidth]{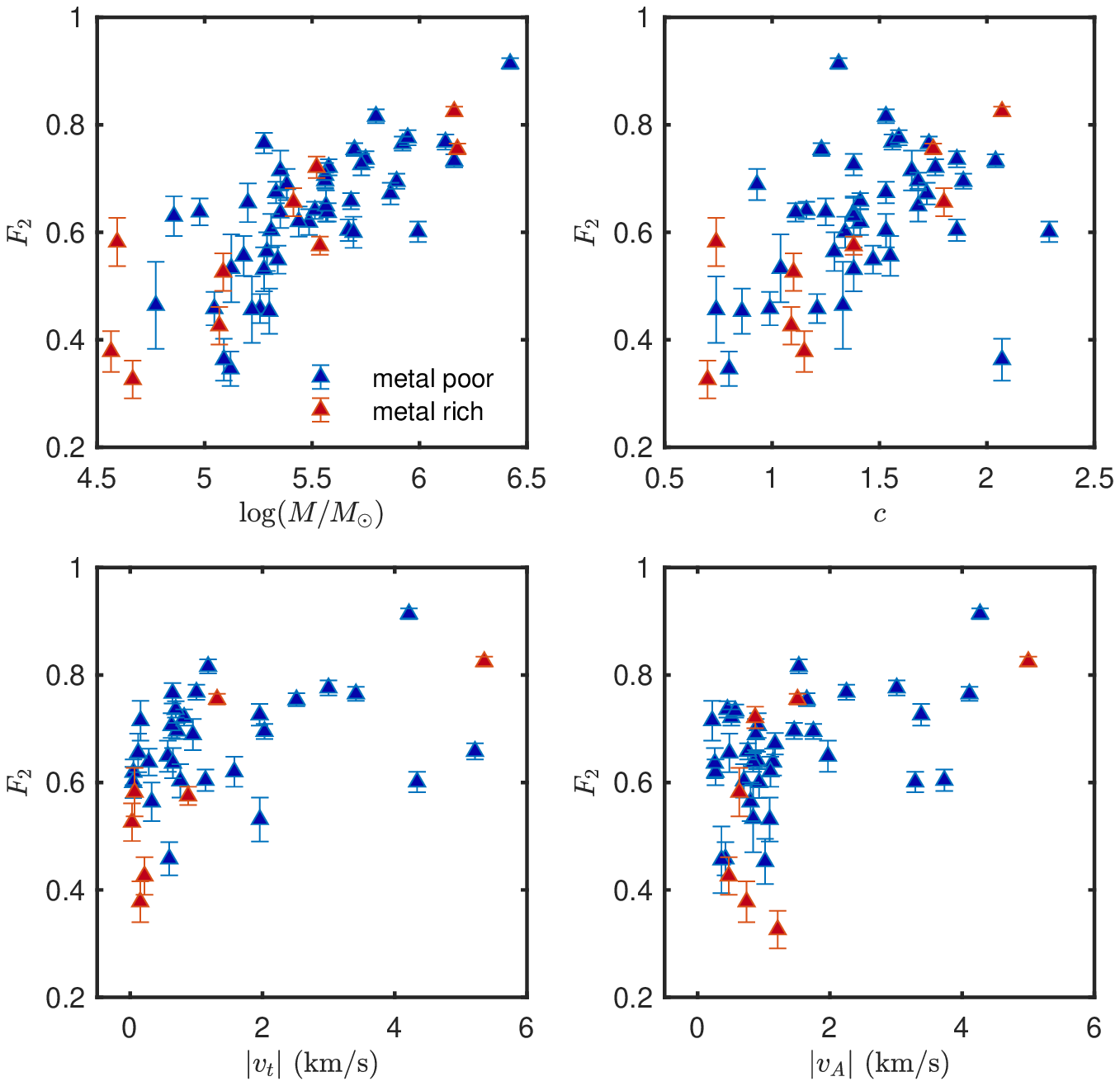}
\caption{$F_2$ is plotted against the same parameters as in Figure \ref{fig:strongcorr}. Blue triangles are metal poor clusters while red triangles  are are metal rich clusters.}\label{fig:appen_1}
\end{figure*}

We tested the same correlations studied in Sect. \ref{sec:res} dividing the clusters into two groups depending on their metallicities instead of their masses.
Low metallicity clusters have $[Fe/H]<-1$, high metallicity clusters have $[Fe/H]\geq -1$.\\
The results of this analysis are presented in Figures \ref{fig:appen_1} and in Tables \ref{tab:app_tab1}-\ref{tab:app_tab2}. 
The differences in the statistical parameters obtained for the entire sample of GCs  when using either the masses or the metallicities are due to the fact that these quantities are not available for the same clusters. 
We do not observe any strong trend of the correlations with the metallicity.\\
Most of the correlations  remain valid (or absent) for both groups when splitting the GCs according to their metallicities. The amplitude of the rotational velocity is correlated to $F_2$ only when considering the entire GC sample and for metal-poor clusters.  This might be linked to the conditions at formation for in-situ clusters \citep[see e.g.][]{Piatti20} or, more likely, simply be caused by the low number of clusters belonging to the metal rich subgroup.
The dynamical age of the clusters seems correlated to $F_2$ only in the case of the Pearson statistical test and when considering all the clusters and metal-poor clusters only.\\
 We detect a slight anticorrelation between $F_2$ and the number of orbits travelled by the clusters, ${ age}/T_r$: the Kendal $p_{\rm value}$ is just above $0.05$ for the metal-rich clusters.  \\
Interestingly, metal-rich clusters sit at the edge of the metal-poor cluster population in the $F_2$-dynamical age plane.\\
The $V/\sigma$ parameter is not correlated to $F_2$ when separating them in two metallicity sub-groups.\\
 The concentration and the mass are strongly correlated  for the entire sample of GCs and they are less strongly correlated to the 2P fraction for the metal rich clusters than for the metal-poor GCs.  The mass and the concentration are mutually correlated for the entire cluster sample and the metal poor clusters. In the case of the metal rich clusters these quantities are, instead, correlated only in the case of the Kendal statistics.

%%%%%%%%%%%%%%%%%%%%%%%
\begin{table*}\centering
\small
\begin{tabular}{ccccccccc}
\hline
Par$_{1}$ & Par$_{2}$ & Stat & PCC & PP & PCC$_{MP}$ & PP$_{MP}$ &PCC$_{MR}$ & PP$_{MR}$\tabularnewline
\hline 
$F_{2}$ & $M$ & P &0.76 & $3.2\times10^{-11}$  &  0.70 & $1.0\times10^{-7}$ &  0.86 & 0.0016 \tabularnewline
        &     & S &0.77 &$1.5\times10^{-11}$ & 0.70 & $1.1\times10^{-7}$& 0.78  & $0.012$  \tabularnewline
        &     & K &0.58 &$4.9\times10^{-10}$ & 0.53 &$4.0\times10^{-7}$& 0.64  & 0.0091  \tabularnewline [0.1cm]
        
  $F_{2}$ & $c$  & P &0.50& $2.9\times10^{-4}$ &0.35& $0.027$ &0.83& $0.0059$\tabularnewline
 & & S & 0.55  & $4.5\times10^{-5}$ & 0.41& $0.0089$ & 0.77 & $0.021$\tabularnewline
 &  & K & 0.40  & $6.2\times10^{-5}$ & 0.30& $0.0078$ & 0.61 & $0.025$\tabularnewline [0.1cm]

$F_{2}$ & $|v_{t}|$ & P &0.46& $0.0044$ &  0.31 & $0.10$ & 0.78 &$0.037$\tabularnewline
 &  &  S &0.54 & $8.7\times10^{-4}$ & $0.35$ & $0.060$ & 0.61 & 0.17\tabularnewline
 &  & K &  0.38 &$8.0\times10^{-4}$ & $0.26$ & $0.053$ & 0.52 & 0.13\tabularnewline [0.1cm]

$F_{2}$ & $|v_{A}|$ & P &  0.47& $0.0016$ & $0.45$ & 0.0064&  0.61 & $0.15$\tabularnewline
 &  & S &0.43& $0.0042$ & $0.36$ & $0.033$ &$0.54$ & 0.24  \tabularnewline
 &  & K &0.30& $0.0052$ & $0.26$ & $0.032$ &0.43 & 0.24  \tabularnewline [0.1cm]

$F_{2}$ & $V/\sigma$ & P &0.28& $0.11$ & $0.052$ &  0.79 &0.74 & 0.091 \tabularnewline
 & & S & $0.33$  & $ 0.056$ & $0.099$& 0.61 &0.60 & 0.24 \tabularnewline
 &  & K & $0.21$ & $ 0.083$ & $0.057$& 0.68 &0.47 & 0.27 \tabularnewline [0.1cm]

$F_{2}$ & $\varepsilon$ & P &0.082& $0.57$ &0.18& $0.27$ & $-0.048$ &$0.90$\tabularnewline
 &  & S & $0.079$ & $0.58$ & $0.14$ & 0.38 & 0.055 & 0.88\tabularnewline
 &  & K & $0.058$ & $0.57$ & $0.10$ & 0.36 & 0.046 & 0.93\tabularnewline [0.1cm]
        
$F_{2}$ & $age$ &  P &0.0035 & $0.98$ & $-0.21$& $0.20$ &0.19& $0.63$\tabularnewline
 &  & S& $-0.050$   & 0.74 & $-0.23$ & 0.14 &0.53 &  0.15 \tabularnewline
 & & K &  $-0.39$   & 0.72 & $-0.16$ & 0.16 &0.51 &  0.084 \tabularnewline [0.1cm]
 
$F_{2}$ & $age/t_{\rm rh}$ & P  & $-0.31$ & $0.031$ &$-0.35$& $ 0.029$ & $-0.064$ & $0.87$\tabularnewline
 &  & S &  $ -0.17$ & $0.26$ & $-0.075$ & 0.65& $0.083$ & 0.84\tabularnewline
 &  & K & $ -0.11$ & $0.27$ & $-0.048$ & 0.67 & 0.17 & 0.61\tabularnewline [0.1cm]

$F_{2}$ & $\log(t_{\rm rh})$ & P & 0.25 & $0.071$ & 0.22 & $0.15$ &0.20& $0.57$\tabularnewline
 &  & S &0.080& $0.61$ & $-0.018$ & $0.97$ & $0.14$ &0.31\tabularnewline
 &  & K &$0.093$& $0.32$ & $0.047$ & 0.66& $-0.067$ &0.86\tabularnewline [0.1cm]
 
$F_2$ & $age/T_r$ & P & $-0.025$ & $0.88$ & $-0.031$ & $0.87$ & $0.61$ & $0.20$ \tabularnewline 
 &  & S & $-0.021$ & $0.90$ & $-0.011$ & $0.96$ & $0.83$ & $0.58$ 
 \tabularnewline
 &  & K & $-0.015$ & $0.91$ & $-0.011$ & $0.95$ & $0.73$ & $0.056$ \tabularnewline
\hline 
\end{tabular}
\caption{Statistical correlation coefficients (PCC) and probabilities (PP) between $F_2$ and several GC internal parameters. Parameters are reported for the entire sample of GCs, for metal poor (MP) and metal rich (MR) clusters. P stands for Pearson, S for Spearman and K for Kendall. 
}\label{tab:app_tab1}
\end{table*}
%%%%%%%%%%%%%%%%%%%%%%%

\begin{table*}\centering
\small
\begin{tabular}{ccccccccc}
\hline
Par$_{1}$ & Par$_{2}$ & Stat & PCC & PP & PCC$_{MP}$ & PP$_{MP}$ &PCC$_{MR}$ & PP$_{MR}$\tabularnewline
\hline 

$|v_{A}|$ & $M$ & P &0.47& $1.5\times10^{-4}$& 0.45 & $0.0014$ &0.53& $0.050$ \tabularnewline
 & & S &0.51& $3.1\times10^{-5}$& $0.51$& $2.4\times10^{-4}$ & $0.60$ & 0.026 \tabularnewline
 & & K &$0.35$& $8.2\times10^{-5}$& $0.36$& $3.2\times10^{-4}$ & $0.41$ & 0.047 \tabularnewline [0.1cm]

$|v_{A}|$ &  c  & P &0.41 & $0.0017$ & 0.39 & $0.0092$ & 0.45& $0.11$\tabularnewline
 &    & S & $0.40$ & $0.0023$ & $0.34$ & 0.025 & 0.62 &  0.018 \tabularnewline
 &    & K & $0.27$ & $0.0033$ & $0.25$ & 0.021 & 0.41 &  0.048 \tabularnewline [0.1cm]

$|v_{A}|$ & age  & P &$0.049$& $0.73$ & 0.11 & $0.49$ &$-0.092$ & $0.80$\tabularnewline
 &   & S & $0.18$ & $0.25$ & $-0.57$ & 0.089 & $0.065$ & 0.64 \tabularnewline
 &   & K & $-0.055$ & $0.61$ & $0.054$ & 0.66 & $-0.32$ & 0.30 \tabularnewline [0.1cm]

$|v_{A}|$ & $age/t_{\rm rh}$ & P  & $-0.28$ &0.040 & $ -0.23$ &0.13 & $-0.46$& 0.19\tabularnewline
 &  & S & $-0.29$ &0.034 &  $-0.25$ &0.097 & $-0.45$& 0.19 \tabularnewline
 &  & K & $-0.20$ &0.028 &  -0.18 &0.086 & $-0.42$& 0.11 \tabularnewline [0.1cm]

$|v_{A}|$ & $\varepsilon$  & P & 0.22& $0.11$ &  0.26 & $0.10$ & 0.052 & $0.86$\tabularnewline
 &   & S & $0.14$ & $0.32$ & $0.18$ & 0.28 & $-0.0061$ & 0.99 \tabularnewline
 &   & K & $0.19$ & $0.045$ & $0.24$ & 0.028 & 0.10& 0.66 \tabularnewline [0.1cm]

$M$ & $V/\sigma$ & P  &0.16& $0.28$ & 0.078& $0.64$ &$0.61$& $0.083$\tabularnewline
 &  & S & $ 0.34$ & $ 0.019$ & $0.20$ & 0.22 & 0.75 & 0.026\tabularnewline
 &  & K & $ 0.21$ & $ 0.037$ & 0.11&  $0.31$ & 0.56 & 0.045 \tabularnewline [0.1cm]

$M$ &  $c$   & P & $ 0.30$ & $0.0021$ & 0.33&  $0.0031$ &0.25 & 0.22\tabularnewline
 &     & S & $ 0.47$ & $5.3\times10^{-7}$ & 0.51& $1.5\times10^{-6}$&  0.32 & 0.11\tabularnewline
 &     & K & $ 0.34$ & $ 3.6\times10^{-7}$ & 0.36&  $ 3.1\times10^{-6}$ &0.29 & 0.042\tabularnewline [0.1cm]

$M$ &  $age$   & P & $ 0.010$ & $0.94$ &$ -0.27$&  $0.090$ &0.35 & 0.27\tabularnewline
 &     & S & $ 0.030$ & $0.83$ & $-0.21$& $0.17$&  0.39 & 0.21\tabularnewline
 &     & K & $8.8\times10^{-3}$ & $0.93$ & $-0.15$&  $0.17$ & 0.28 & 0.26\tabularnewline [0.1cm]

$M$ & $age/t_{\rm rh}$ & P  &$-0.31$ &0.019 & $ -0.32$ &0.042 & $ -0.29$& 0.37 \tabularnewline
 &   & S &$-0.27$ &0.37 &  $-0.26$ &0.097 & $-0.049$& 0.89 \tabularnewline
 &   & K &$-0.20$ & 0.036 & $ -0.19$ &0.081 &$-0.030$ &0.95 \tabularnewline [0.1cm]

$M$ & $\varepsilon$  & P & $ 0.17$ & $ 0.096$ & 0.065&  $0.64$ &0.28 & 0.064 \tabularnewline 
 &  & S & $ 0.17$ & $ 0.10$ & 0.15&  $0.19$ &0.14 & 0.52 \tabularnewline 
 &  & K & $ 0.13$ & $ 0.076$ & $0.11$&  $0.16$ &0.14 & 0.34 \tabularnewline [0.1cm]
\hline 
\end{tabular}
\caption{Statistical correlation coefficients (PCC) and probabilities (PP) between the rotational velocity $v_A$, age and several GC internal parameters. Parameters are reported for the entire sample of GCs, for metal poor (MP) and metal rich (MR) clusters. P stands for Pearson, S for Spearman and K for Kendall. 
}
\label{tab:app_tab2}
\end{table*}

%%%%%%%%%%%%%%%%%%%%%%

%%%%%%%%%%%%%%%%%%%%%%%%%%%%%%%%%%%%%%

\begin{figure*}\centering
\includegraphics[width=0.85\textwidth]{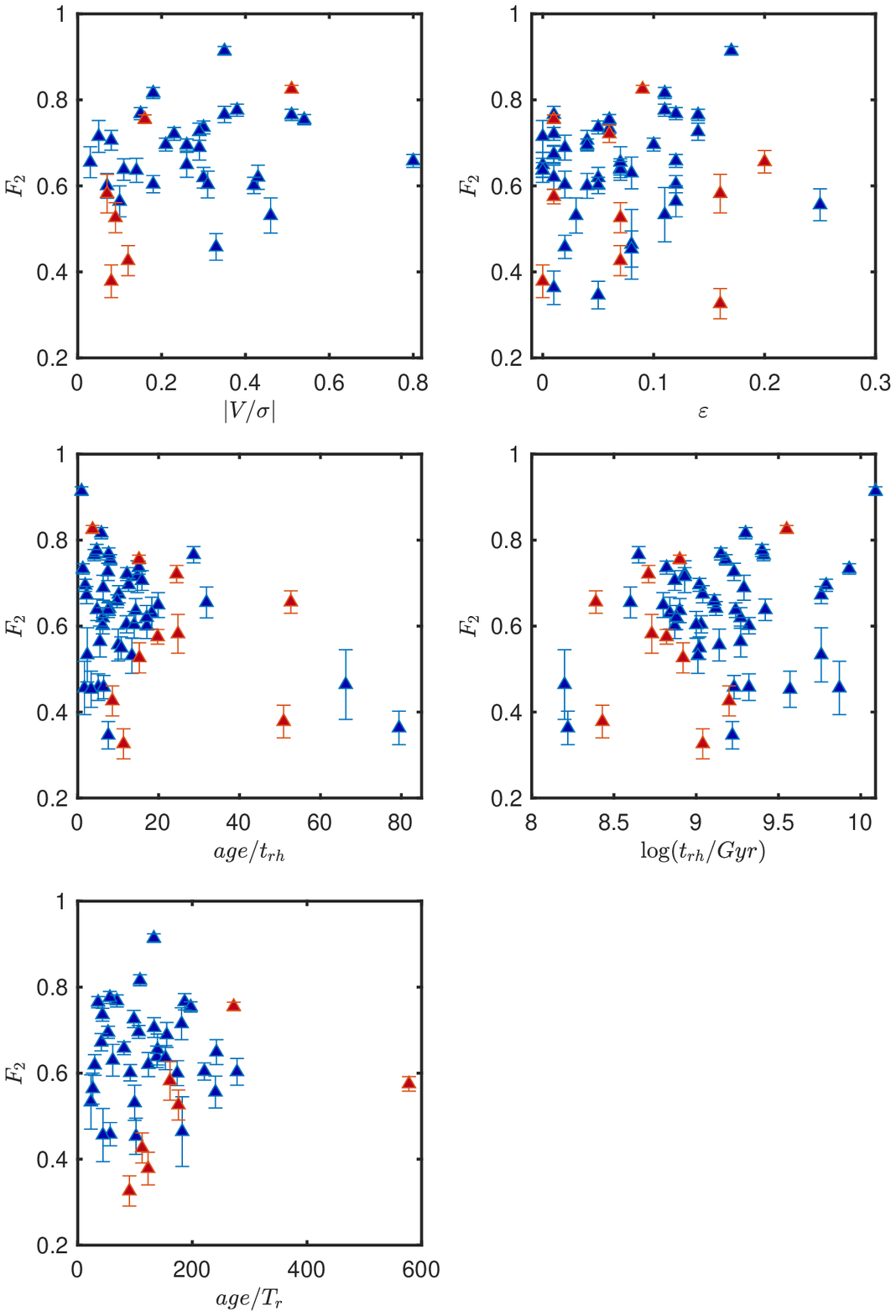}
\caption{$F_2$ is plotted against the same parameters as in Figure \ref{fig:other_corr}. Blue triangles are metal poor clusters while red triangles  are are metal rich clusters.}\label{fig:appen_2}
\end{figure*}

%\begin{figure*}\centering
%\includegraphics[width=0.85\textwidth]{FigureA3.eps}
%\caption{$v_A$ is plotted against the same parameters as in Figure %\ref{fig:vtcorr}. Blue triangles are metal poor clusters while red %triangles  are are metal rich clusters.}\label{fig:appen_3}
%\end{figure*}

%\begin{figure*}\centering
%\includegraphics[width=0.85\textwidth]{FigureA4.eps}
%\caption{The cluster mass, $M$, is plotted against the same %parameters as in Figure \ref{fig:mass}. Blue triangles are metal poor %clusters while red triangles  are are metal rich %clusters.}\label{fig:appen_4}
%\end{figure*}

\section{Correlations and projected rotational velocities}\label{sec:app2}

For completeness, we have also tested the projected rotational velocity, $v_t$, against the other internal parameters considered in this work. The results obtained considering the metallicity threshold are reported in %Figure \ref{fig:appen_5} and 
Table \ref{tab:app_tab3}, while the results obtained considering the mass threshold are presented in
%Figure \ref{fig:appen_6} and 
Table \ref{tab:app_tab4}.

%%%%%%%%%%%%%%%%%%%%%%
\begin{table*}\centering
\small
\begin{tabular}{ccccccccc}
\hline
Par$_{1}$ & Par$_{2}$ & Stat & PCC & PP & PCC$_{MP}$ & PP$_{MP}$ &PCC$_{MR}$ & PP$_{MR}$\tabularnewline
\hline 

$|v_{t}|$ & $M$ & P &0.58& $6.9\times10^{-6}$& 0.16 & $0.48$ &0.42& $0.024$ \tabularnewline
 & & S &0.64& $3.6\times10^{-7}$& $0.18$& $0.44$ & $0.49$ & 0.0070 \tabularnewline
 & & K &$0.46$& $1.7\times10^{-6}$& $0.13$& $0.43$ & $0.33$ & 0.012 \tabularnewline [0.1cm]

$|v_{t}|$ &  c  & P &0.40 & $0.0068$ & 0.20 & $0.43$ & 0.22& $0.28$\tabularnewline
 &    & S & $0.40$ & $0.0065$ & $0.23$ & 0.37 & 0.091 &  0.65 \tabularnewline
 &    & K & $0.28$ & $0.0071$ & $0.18$ & 0.32 & 0.060 &  0.68 \tabularnewline [0.1cm]

$|v_{t}|$ & $age$  & P &$0.17$& $0.32$ & 0.094 & $0.64$ &$0.29$ & $0.49$\tabularnewline
 &   & S & $0.14$ & $0.42$ & $-0.011$ & 0.96 & $0.29$ & 0.48 \tabularnewline
 &   & K & $0.092$ & $0.47$ & $-0.018$ & 0.92 & $0.23$ & 0.53 \tabularnewline [0.1cm]

$|v_{t}|$ & $age/t_{\rm rh}$ & P  &-0.43 &$0.010$ &  $-0.37$ &0.058 & $-0.50$& 0.20\tabularnewline
 &  & S &-0.52 & 0.0015 &  $-0.41$ &0.034 & $-0.52$& 0.20 \tabularnewline
 &  & K &-0.37 &$0.0019$ &  $-0.31$ &0.023 & $-0.36$& $0.28$ \tabularnewline [0.1cm]

$|v_{t}|$ & $\varepsilon$  & P & $0.14$& $0.32$ &  $-0.39$ & $0.079$ & 0.22 & $0.24$\tabularnewline
 &   & S & $0.14$ & $0.32$ & $0.18$ & 0.28 & $-0.0061$ & 0.99 \tabularnewline
 &   & K & $0.10$ & $0.31$ & $-0.24$ & 0.14 & 0.16& 0.26 \tabularnewline [0.1cm]

\hline 
\end{tabular}
\caption{Statistical correlation coefficients (PCC) and probabilities (PP) between the projected rotational velocity, $v_t$, and several GC internal parameters. Parameters are reported for the entire sample of GCs, for metal-poor (MP) and metal-rich (MR) clusters. P stands for Pearson, S for Spearman and K for Kendall. 
}
\label{tab:app_tab3}
\end{table*}
%%%%%%%%%%%%%%%%%%%%%%%%%%%%%%%%%%%%%%
\begin{table*}\centering
\small
\begin{tabular}{ccccccccc}
\hline
Par$_{1}$ & Par$_{2}$ & Stat & PCC & PP & PCC$_{LM}$ & PP$_{LM}$ &PCC$_{HM}$ & PP$_{HM}$\tabularnewline
\hline 

$|v_{t}|$ & $M$ & P &0.58& $6.9\times10^{-6}$& 0.52 & $4.4\times10^{-4}$ &0/77& $0.0097$ \tabularnewline
 & & S &0.64& $3.6\times10^{-7}$& $0.58$& $7.8\times10^{-5}$ & $0.83$ & 0.0056 \tabularnewline
 & & K &$0.46$& $1.7\times10^{-6}$& $0.40$& $2.1\times10^{-4}$ & $0.64$ & 0.0091 \tabularnewline [0.1cm]

$|v_{t}|$ &  c  & P &0.40 & $0.0068$ & 0.26 & $0.13$ & 0.70& $0.022$\tabularnewline
 &    & S & $0.40$ & $0.0065$ & $0.25$ & 0.14 & 0.82 &  0.0068 \tabularnewline
 &    & K & $0.28$ & $0.0071$ & $0.18$ & 0.13 & 0.69 &  0.0047 \tabularnewline [0.1cm]

$|v_{t}|$ & $age$  & P &$0.17$& $0.32$ & 0.021 & $0.18$ &$-0.47$ & $0.61$\tabularnewline
 &   & S & $0.14$ & $0.42$ & $0.54$ & 0.024 & $0.039$ & 0.88 \tabularnewline
 &   & K & $0.092$ & $0.47$ & $0.44$ & 0.018 & $0.021$ & 0.94 \tabularnewline [0.1cm]

$|v_{t}|$ & $age/t_{\rm rh}$ & P  & $-0.43$ &$0.010$ &  $-0.13 $&0.61 & $-0.66$& 0.0032\tabularnewline
 &  & S &$-0.52$ &$0.0015$ &  $-0.12$ &0.64 & $-0.82$& $3.1\times10^{-5}$ \tabularnewline
 &  & K &$-0.37$ &0.0019 &  $-0.074$ &0.72 & $-0.66$& $1.8\times10^{-4}$ \tabularnewline [0.1cm]

$|v_{t}|$ & $\varepsilon$  & P & 0.17& $0.096$ &  0.19 & $0.11$ & 0.20 & $0.35$\tabularnewline
 &   & S & $0.14$ & $0.32$ & $0.18$ & 0.28 & $-0.0061$ & 0.99 \tabularnewline
 &   & K & $0.10$ & $0.30$ & $0.13$ & 0.26 & 0.023& 1.0 \tabularnewline [0.1cm]

\hline 
\end{tabular}
\caption{Statistical correlation coefficients (PCC) and probabilities (PP) between the projected rotational velocity, $v_t$, and several GC internal parameters. Parameters are reported for the entire sample of GCs, for clusters more massive (HM) and less massive (LM) than $10^{5.5}~M_\odot$. P stands for Pearson, S for Spearman and K for Kendall.
}
\label{tab:app_tab4}
\end{table*}
%%%%%%%%%%%%%%%%%%%%%%%%%%%%%%%%%%%%%%
\begin{table*}\centering
\small
\begin{tabular}{ccccccccc}
\hline
Par$_{1}$ & Par$_{2}$ & Stat & PCC & PP & PCC$_{LM}$ & PP$_{LM}$ &PCC$_{HM}$ & PP$_{HM}$\tabularnewline
\hline 

$F_2$& $f(r<r_c)$ & P &$-0.39$& $0.024$& $-0.12$ & $0.57$ & $-0.89$& $0.0033$ \tabularnewline
 & & S &$-0.49$& $0.0032$& $-0.20$ & $0.33$ & $-0.92$& $0.0025$ 
 \tabularnewline
 & & K &$-0.38$& $0.0017$& $-0.15$ & $0.28$ & $-0.84$& $0.0028$ 
 \tabularnewline [0.1cm]

$F_2$& $f(r_c\leq r < r_h)$ &  P &$-0.72$ & $2.8\times10^{-9}$ & $-0.44$ & $0.018$ & $-0.51$ & $0.015$\tabularnewline
 &    & S &$-0.58$ & $92.6\times10^{-5}$ & $-0.31$ & $0.12$ & $-0.41$ & $0.091$ \tabularnewline
 &    & K &$-0.41$ & $9.7\times10^{-5}$ & $-0.20$ & $0.16$ & $-0.29$ & $0.10$\tabularnewline [0.1cm]

$F_2$& $f(r \geq r_h)$  & P &$-0.69$& $1.3\times10^{-6}$ & $0.59$ & $0.0064$ &$-0.30$ & $0.21$\tabularnewline
 &   & S &$-0.73$& $1.7\times10^{-7}$ & $0.66$ & $0.0014$ &$-0.40$ & $0.091$ \tabularnewline
 &   & K &$0.55$& $9.3\times10^{-7}$ & $0.49$ & $0.0031$ &$-0.28$ & $0.11$
 \tabularnewline [0.1cm]

$F_2$& $f_{WFC}$ & P  & $-0.68$ &$5.6\times10^{-8}$ &  $-0.46 $&0.014 & $-0.56$& 0.0068\tabularnewline
 &  & S & $-0.72$ &$2.8\times10^{-9}$ &  $-0.44 $&0.018 & $-0.51$& 0.015
 \tabularnewline
 &  & K & $-0.53$ &$3.4\times10^{-8}$ &  $-0.31 $&0.022 & $-0.30$& 0.049 \tabularnewline [0.1cm]

$\log(M)$& $f_{WFC}$ &  P & $-0.65$& $1.4\times10^{-7}$ &  $-0.46$ & $0.014$ & $-0.40$ & $0.053$\tabularnewline
 &   & S & $-0.73$& $8.5\times10^{-10}$ &  $-0.44$ & $0.018$ & $-0.45$ & $0.026$ \tabularnewline
 &   & K & $-0.73$& $8.5\times10^{-10}$ &  $-0.44$ & $0.018$ & $-0.45$ & $0.026$ \tabularnewline [0.1cm]

\hline 
\end{tabular}
\caption{Statistical correlation coefficients (PCC) and probabilities (PP) between $F_2$ total binary fractions within the core radius ($r_c$), between $r_c$ and the half-mass radius ($r_h$) of the clusters, outside $r_h$ and in the full HST WFC field. Parameters are reported for the entire sample of GCs, for clusters more massive (HM) and less massive (LM) than $10^{5.5}~M_\odot$. P stands for Pearson, S for Spearman and K for Kendall.
}
\label{tab:tab_app_bin}
\end{table*}
%\begin{figure*}\centering
%\includegraphics[width=0.85\textwidth]{FigureA5.eps}
%\caption{The analogue of Figure \ref{fig:appen_3} but for the %projected rotational velocity, $v_t$, of the %clusters.}\label{fig:appen_5}
%\end{figure*}

%\begin{figure*}\centering
%\includegraphics[width=0.85\textwidth]{FigureA6.eps}
%\caption{The analogue of Figure \ref{fig:appen_5} done applying a %ass-based threshold to divide the clusters in two subgroups. %}\label{fig:appen_6}
%\end{figure*}

\section{Age correlations }
\label{app:age}
 
We used four different data-sets to check for the existence of a relationship between $F_2$ and the age of the clusters \citep{FO10, Do10, Gratton10, VDB13}. 
We found a clear anti-correlation between the two quantities only when using the ages taken from  \cite{FO10}. 
In this case, clusters less massive than $10^{5.5}~M_\odot$ do not show any clear $F_2$-age correlation, while clusters more massive than the threshold display a clear trend between the two quantities when outliers are removed.\\
The ages from \cite{FO10} have been, however, collected from several heterogeneous sources and, therefore, the correlation found might be spurious.
Indeed, when using the ages collected in more homogeneous ways \citep{Do10, Gratton10, VDB13}, we do not find any strong relationship between $F_2$ and the cluster age. This is in line with what found by \cite{MI17}, who only detected a mild anti-correlation between the 1P fractions and with the ages from \cite{VDB13}.\\
We note that, when using the same clusters available with \cite{Do10}, \cite{Gratton10} and \cite{VDB13}, we do not find any anti-correlation also when using the ages from \cite{FO10}. Homogeneous and more numerous information on the ages of the clusters are necessary to determine if an anti-correlation between $F_2$ and the cluster age exists or not.

\section{Correlations and binary fractions} \label{sec:app_bin_frac}
 
\begin{figure*}\centering
\includegraphics[width=0.85\textwidth]{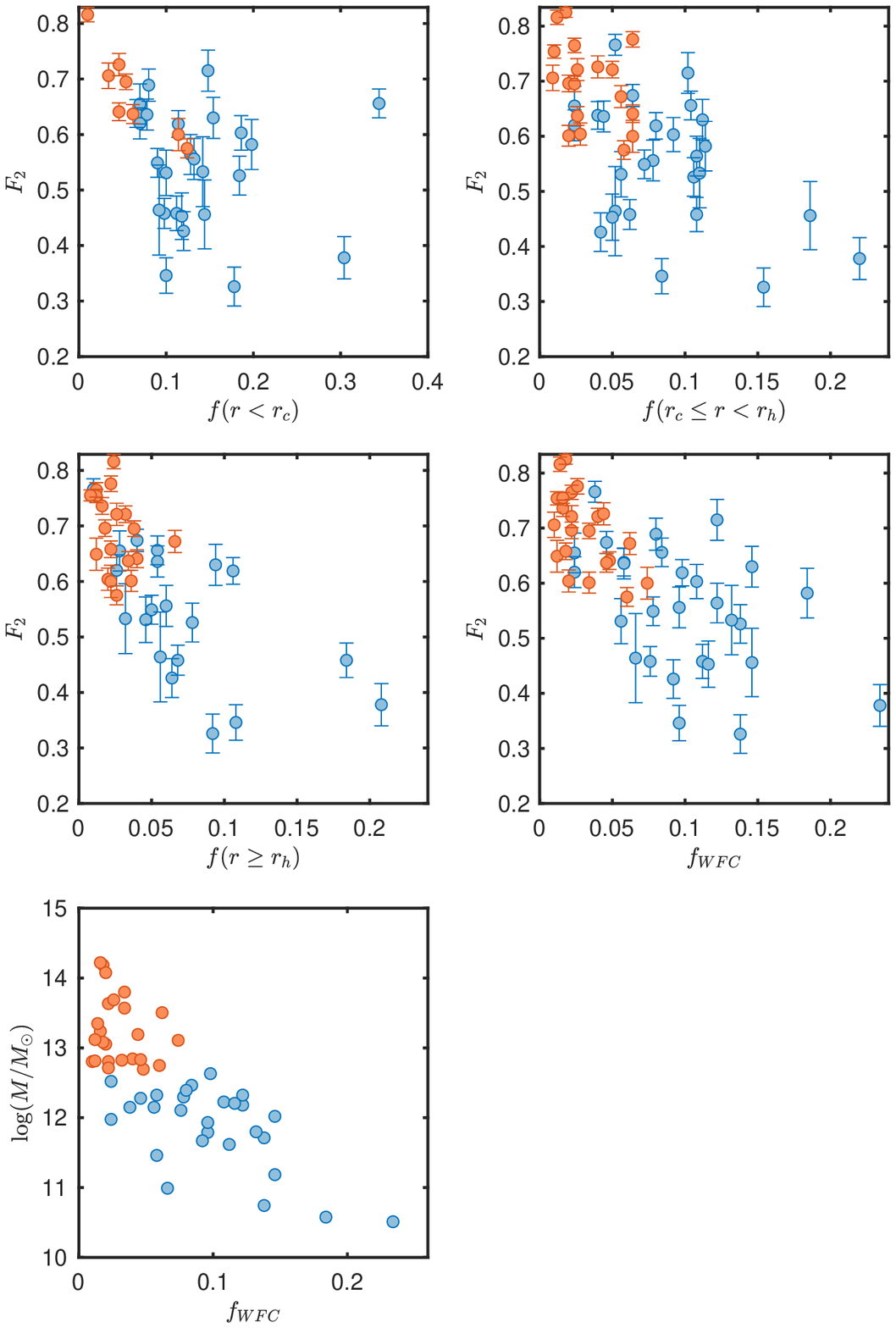}
\caption{Going from the top to the bottom and from left to right, the fraction of 2P stars, $F_2$, is plotted against the binary fraction within the core radius, $f(r<r_c)$, between $r_c$ and the half-mass radius, $f(r_c\leq r<r_h$), outside $r_h$, $f(r\geq r_h)$ and in the full HST WFC field, $f_{WFC}$. The bottom left panel shows the logarithm of the cluster masses as a function of the binary fraction in the entire WFC field. }\label{fig:app_bin}
\end{figure*}

In addition to  the correlations studied above, we have also investigated the existence of a trend between $F_2$ and the cluster binary fractions reported by \cite{Mil12} and \cite{Milone16}. While \cite{Gratton19} only considered the total binary fraction observed in the entire HST WFC field of view,  we have looked for correlations in different radial bins, i.e. within the core radius ($r_c$) of the clusters, between their $r_c$ and half-mass radius  ($r_h$), for radii larger than $r_h$ and in the whole WFC field. 
As \cite{Gratton19}, we find a very strong anticorrelation between $F_2$ and the binary fraction (see Fig. \ref{fig:app_bin} and Table \ref{tab:tab_app_bin}). The anticorrelation is the tightest when considering the whole field of view ($f_{WFC}$) and its strength decreases when we consider more internal radial bins. When looking at distances smaller than $r_h$, the anticorrelation is stronger for clusters more massive than $10^{5.5~M_\odot}$ . At larger radii, as well as when we consider the entire WFC field, the anticorrelation is, instead, more pronounced for low mass clusters.\\
Additionally, we observe an extremely thigh anticorrelation between the mass of the clusters and $f_{WFC}$. Thus, we repeated the same regression analysis performed in Sect. \ref{sec:majorcorr}, this time considering $f_{WFC}$ as an additional parameter. The result of the regression confirms that the mass is the most important parameter ($p_{\rm value}=9.8\times10^{-5}$) driving the 2P phenomenon, followed by  $f_{\rm WFC}$ ($p_{\rm value}=0.079$). As suggested by \cite{Gratton19}, the anticorrelation between the binary fraction and $F_2$ could be evidence of the fact that 2P are born more concentrated and, therefore, with a smaller binary fraction due to a higher rate of interactions in the inner cluster regions. However, in this case, the anticorrelation should become stronger in more internal bins. The fact that we observe the exact opposite might be due to observational difficulties and to the lower number of binaries available in more collisional regions or imply that the relationship between $F_2$ and the binary fraction might also be mediated by the mass.

% Don't change these lines
\bsp	% typesetting comment
\label{lastpage}
\end{document}